\pdfoutput=1
\documentclass[useAMS,usenatbib]{mn2e}
\usepackage{apjfonts}
\usepackage{graphicx}
\usepackage{amsmath}
\usepackage{amssymb}
\usepackage{ctable}
\usepackage{fixltx2e}
\usepackage{threeparttable}
\usepackage{hyperref}
\hypersetup{colorlinks=true,linkcolor=blue,citecolor=blue,filecolor=blue,urlcolor=blue}

\newcommand{\be}{\begin{equation}}
\newcommand{\ee}{\end{equation}}
\newcommand{\ba}{\begin{eqnarray}}
\newcommand{\ea}{\end{eqnarray}}

\newcommand{\Rvir}{R_{\rm vir}}

\newcommand{\Mhalo}{M_{\rm halo}}
\newcommand{\Rmax}{R_{\rm max}}

\newcommand{\Ms}{M_{\ast}}

\newcommand{\SFR}{{\rm SFR}}
\newcommand{\cm}{{\rm cm}}

\newcommand{\pc}{{\rm pc}}

\newcommand{\Mpc}{{\rm Mpc}}
\newcommand{\dex}{{\rm dex}}

\newcommand{\g}{{\rm g}}
\newcommand{\yr}{{\rm yr}}
\newcommand{\Myr}{{\rm Myr}}

\newcommand{\Msun}{M_{\sun}}
\newcommand{\Lsun}{L_{\sun}}
\newcommand{\Zsun}{Z_{\sun}}

\newcommand{\HII}{H\,{\sc ii}}

\newcommand{\mb}{m_b}
\newcommand{\mDM}{m_{\rm DM}}
\newcommand{\eg}{\epsilon_{\rm gas}}
\newcommand{\es}{\epsilon_{\rm star}}
\newcommand{\eDM}{\epsilon_{\rm DM}}
\newcommand{\hinv}{h^{-1}}
\newcommand{\nc}{n_{\rm th}}

\newcommand{\fdust}{f_{\rm dust}}
\newcommand{\kdust}{\kappa_{\rm dust}}
\newcommand{\Mdust}{M_{\rm dust}}
\newcommand{\Lbol}{L_{\rm bol}}
\newcommand{\LUV}{L_{\rm UV}}
\newcommand{\LUVi}{L_{\rm UV,\,intr}}
\newcommand{\LIR}{L_{\rm IR}}
\newcommand{\FUV}{F_{\rm UV}}
\newcommand{\FIR}{F_{\rm IR}}
\newcommand{\Teff}{T_{\rm eff}}
\newcommand{\IRX}{{\rm IRX}}
\newcommand{\buv}{\beta_{\rm UV}}
\newcommand{\muv}{{\rm M_{UV}}}
\newcommand{\lp}{\lambda_{\rm peak}}
\newcommand{\Tp}{T_{\rm peak}}
\newcommand{\dd}{{\rm d}}
\newcommand{\LCDM}{$\Lambda$CDM}

\newcommand{\apj}{ApJ}
\newcommand{\apjl}{ApJ}
\newcommand{\apjs}{ApJS}
\newcommand{\mnras}{MNRAS}
\newcommand{\aap}{A\&A}
\newcommand{\araa}{ARA\&A}
\newcommand{\nat}{Nature}
\newcommand{\rmxaa}{RMxAA}

\newcommand{\pasj}{PASJ}
\newcommand{\pasa}{PASA}

\title[Dust in high-$z$ galaxies]
{Dust attenuation, dust emission, and dust temperature in galaxies at $z\geq5$: a view from the FIRE-2 simulations}

\author[X. Ma et al.]{
  \parbox[t]{1.0\textwidth}{
   Xiangcheng Ma,$^{1,2}$\thanks{E-mail: xchma@berkeley.edu}
   Christopher C. Hayward,$^3$
   Caitlin M. Casey,$^4$
   Philip F. Hopkins,$^2$ \\
   Eliot Quataert,$^1$
   Lichen Liang,$^5$
   Claude-Andr{\'e} Faucher-Gigu{\`e}re,$^6$ \\
   Robert Feldmann$^5$ and
   Du{\v s}an Kere{\v s}$^7$
  }
  \vspace{5pt} \\
  $^1$Department of Astronomy and Theoretical Astrophysics Center, University of California Berkeley, Berkeley, CA 94720 \\
  $^2$TAPIR, MC 350-17, California Institute of Technology, Pasadena, CA 91125, USA \\ 
  $^3$Center for Computational Astrophysics, Flatiron Institute, 162 Fifth Avenue, New York, NY 10010, USA \\
  $^4$Department of Astronomy, The University of Texas at Austin, 2515 Speedway Blvd, Stop C1400, Austin, TX 78712, USA \\
  $^5$Institute for Computational Science, University of Zurich, Zurich CH-8057, Switzerland \\
  $^6$Department of Physics and Astronomy and CIERA, Northwestern University, 2145 Sheridan Road, Evanston, IL 60208, USA \\
  $^7$Department of Physics, Center for Astrophysics and Space Sciences, University of California at San Diego, 9500 Gilman Drive, La Jolla, CA 92093 \\
}

\pagerange{\pageref{firstpage}--\pageref{lastpage}}
\date{Draft version \today}

\begin{document}
\maketitle
\label{firstpage}

\begin{abstract}
We present a suite of 34 high-resolution cosmological zoom-in simulations consisting of thousands of halos up to $\Mhalo\sim10^{12}\,\Msun$ ($\Ms\sim10^{10.5}\,\Msun$) at $z\geq5$ from the Feedback in Realistic Environments project. We post-process our simulations with a three-dimensional Monte Carlo dust radiative transfer code to study dust attenuation, dust emission, and dust temperature within these simulated $z\geq5$ galaxies. Our sample forms a tight correlation between infrared excess ($\IRX\equiv\FIR/\FUV$) and ultraviolet (UV)-continuum slope ($\buv$), despite the patchy, clumpy dust geometry shown in our simulations. We find that the IRX--$\buv$ relation is mainly determined by the shape of the attenuation law and is independent of its normalization (set by the dust-to-gas ratio). The bolometric IR luminosity ($\LIR$) correlates with the {\em intrinsic} UV luminosity and the star formation rate (SFR) averaged over the past 10\,Myr. We predict that at a given $\LIR$, the peak wavelength of the dust spectral energy distributions for $z\geq5$ galaxies is smaller by a factor of 2 (due to higher dust temperatures on average) than at $z=0$. The higher dust temperatures are driven by higher specific SFRs and SFR surface densities with increasing redshift. We derive the galaxy UV luminosity functions (LFs) at $z=5$--10 from our simulations and confirm that a heavy attenuation is required to reproduce the observed bright-end UVLFs. We also predict the IRLFs and UV luminosity densities at $z=5$--10. We discuss the implications of our results on current and future observations probing dust attenuation and emission in $z\geq5$ galaxies.
\end{abstract}

\begin{keywords}
galaxies: evolution -- galaxies: formation -- galaxies: high-redshift -- cosmology: theory -- ISM: dust, extinction
\end{keywords}

\section{Introduction}
\label{sec:intro}
Improving the constraints on the star formation rate density (SFRD) across cosmic time is important for understanding the assembly history of galaxies \citep[see][for a recent review]{Madau:2014}. At $z\geq5$ in particular, the cosmic SFRD directly relates to the number of ionizing photons available from star-forming galaxies for cosmic reionization \citep[dependent upon the escape fraction, e.g.][]{Finkelstein:2012,Kuhlen:2012,Robertson:2013,Robertson:2015}, so understanding the SFRD at $z\geq5$ is also crucial for constraining the reionization history.

It is well known that at $z\lesssim3$, the cosmic SFRD is dominated by dusty star-forming galaxies \citep[DSFGs; e.g.][]{Magnelli:2011,Casey:2012a,Gruppioni:2013}, which have very high star formation rates (SFRs) but are too heavily obscured to be seen in the UV and optical. On the other hand, the cosmic SFRD at $z\geq5$ is only probed in the rest-frame UV up to $z\sim10$ \citep[e.g.][]{Coe:2013,Ellis:2013,Oesch:2013,Oesch:2014,Bouwens:2015,Bouwens:2016a,Finkelstein:2015}. A consensus of the obscured fraction of star formation at these redshifts is not yet in place. The most commonly adopted approach to correct for dust obscuration in UV-selected galaxies at $z\geq5$ is to use the empirical relationship between infrared (IR) excess, $\IRX\equiv\FIR/\FUV$, and the UV-continuum slope, $\buv$ \citep[e.g.][]{Bouwens:2015,Finkelstein:2015}. This so-called IRX--$\buv$ relation was first established in local compact starburst galaxies \citep[e.g.][]{Meurer:1999} and has been confirmed to hold up to $z\sim2$--3 \citep[e.g.][]{Reddy:2006,Reddy:2018,Alvarez-Marquez:2016,Bourne:2017,Fudamoto:2017a,McLure:2018}.

However, at $z\geq5$, it is yet unclear whether the IRX--$\buv$ relation, which is a reflection of the dust attenuation law, still applies. \cite{Capak:2015} and subsequently \cite{Barisic:2017} studied a sample of $z\sim5.5$ galaxies and found that they exhibit a large scatter in the IRX--$\buv$ relation (see also \citealt{Bourne:2017} and \citealt{Fudamoto:2017a} for their $z\sim5$ sample; however, \citealt{Koprowski:2018} reported that $z\sim5$ galaxies are still consistent with the local relation), where some galaxies fall significantly below the IRX--$\buv$ relation derived from a steeper, Small Magellanic Cloud (SMC)-like reddening law. Moreover, \cite{Bouwens:2016} showed that deep 1.2\,mm continuum survey with the Atacama Large Millimeter Array (ALMA) in the Hubble Ultra Deep Field (HUDF) detects much fewer high-redshift sources than what inferred from the IRX--$\buv$ relation using the rest-frame UV slopes and luminosities of galaxies in the same field \citep[see also][]{Dunlop:2017}, where a $T\sim35$\,K modified black-body (MBB) function is assumed for dust emission. This also suggests that the IRX of high-redshift galaxies are well below the SMC IRX--$\buv$ relation.

There are two possible explanations of these results: one physical, one observational. First, it is likely that $z\geq5$ galaxies are more dust poor than low- and intermediate-redshift galaxies at the same stellar mass, because the universe has not allowed sufficient time for dust to grow substantially. On the other hand, it is also possible that dust luminosity is severely underestimated due to the assumed spectral energy distributions (SEDs) of dust emission. As noted by \cite{Bouwens:2016}, the tension can be alleviated if dust temperature reaches as high as 45--50\,K at $z\sim6$ \citep[see also e.g.][]{Faisst:2017}, such that dust is much less luminous at long wavelengths {\em at the same total IR luminosity}. Alternatively, even if the cold dust remains $\sim35$\,K, a moderate fraction of warm dust in the galaxy can dramatically reshape the dust SEDs and reduce the apparent flux density on the Rayleigh--Jeans (R--J) tail at fixed IR luminosity \citep[][]{Casey:2018a}. In either case, fundamentally different dust properties are not necessarily required for high-redshift galaxies. Therefore, it is critical to understand (1) in what conditions the local IRX--$\buv$ relations still hold (or not) and (2) the typical dust temperature and SEDs in galaxies above $z\sim5$ to properly account for obscured star formation from pure rest-frame UV surveys.

Current knowledge on the population of DSFGs at $z\geq5$ is still limited. At the extremely luminous, high-SFR end ($\LIR\sim10^{13}\,\Lsun$), there is a growing sample of DSFGs at $z\sim5$--7 built over the past few years \citep[e.g.][]{Riechers:2013,Vieira:2013,Weis:2013,Strandet:2016,Marrone:2018}. Dust emission has also been detected in a small number of less extreme systems even at higher redshifts (e.g. \citealt{Watson:2015}, $z\sim7.5$; \citealt{Laporte:2017a}, $z\sim8.38$; \citealt{Hashimoto:2018a}, $z\sim7.15$; \citealt{Tamura:2018}, $z\sim8.3$), many of which are gravitationally lensed galaxies. It is obvious from these observational facts that dust plays a non-negligible role in normal star-forming galaxies even in the very early Universe. However, there is still a lack of efficient ways for finding large samples of DSFGs at moderate luminosities (e.g. $\LIR\sim10^{11}$--$10^{12}\,\Lsun$) at $z\sim5$ and beyond.

Future observational facilities have been proposed, including the Chajnantor Sub/Millimeter Survey Telescope \citep[CSST;][]{Golwala:2018}, the next-generation Very Large Array \citep[ngVLA;][]{McKinnon:2016}, the TolTEC camera on the Large Millimeter Telescope \citep[LMT;][]{Bryan:2018}, the Space Infrared telescope for Cosmology and Astrophysics \citep[SPICA;][]{Egami:2018} and the Origins Space Telescope \citep[OST;][]{Battersby:2018}. Together with ALMA, these facilities are expected to advance our knowledge on high-redshift DSFGs in greater detail. To maximize the efficiency of future observations, it is of great importance to make useful predictions on the properties of DSFGs at $z\geq5$. For example, what are the IR luminosity functions (LFs; i.e. the number density of galaxies at a given $\LIR$) at these redshifts \citep[e.g.][]{Casey:2018}? Is it reliable to infer dust luminosities from the UV-continuum slopes (e.g. as a way of finding targets to observe at longer wavelengths)? What wavelengths are most useful to probe to robustly measure $\LIR$ and dust temperature as well as to avoid severe contamination from the Cosmic Microwave Background \cite[CMB;][]{da-Cunha:2013}? Are there independent observables in the UV that can be combined with IR data to better constrain the dust properties at $z\geq5$?

From a theoretical point of view, it has been broadly appreciated that dust attenuation plays a key role in shaping the bright-end UVLFs at $z\geq5$. Numerous studies have demonstrated this both in semi-analytic models \citep[SAMs; e.g.][]{Clay:2015,Liu:2016,Cowley:2018,Tacchella:2018,Yung:2018} and in cosmological simulations \citep[e.g.][]{Cullen:2017,Wilkins:2017,Ma:2018a}. Many of these previous studies have assumed simple prescriptions for dust attenuation. For example, \cite{Ma:2018a} calculated the integrated optical depth along a given sightline from every star particle in the simulations and applied an attenuation $e^{-\tau}$ for individual particles to obtain the post-attenuation UV luminosity \citep[see also e.g.][]{Katz:2018}. Such simple treatment does not properly account for certain radiative transfer effects, such as dust scattering, which can be important in the UV and optical \citep[e.g.][]{Barrow:2017}. Moreover, dust temperature and SEDs cannot be self-consistently calculated from this approach \citep[e.g.][]{Wilkins:2018}, which limits the predictive power of these calculations. To this end, full dust radiative transfer calculations are necessary.

In recent years, post-processing dust radiative transfer calculations have been conducted on large-volume cosmological simulations, cosmological zoom-in simulations, and idealized simulations of disks and mergers to investigate a broad range of questions, including dust attenuation and emission in local galaxies \citep[e.g.][]{Camps:2016,Trayford:2017}, the physical origin of the IRX--$\buv$ relation \citep[e.g.][]{Safarzadeh:2017,Narayanan:2018}, the effects of dust geometry on the reddening law \citep[e.g.][]{Narayanan:2018a}, and the empirical relation between far-IR/submm flux and molecular gas mass \citep[e.g.][]{Liang:2018,Privon:2018}. More relevant to high-redshift galaxies, \cite{Behrens:2018} carried on dust radiative transfer calculation on one galaxy from a high-resolution cosmological zoom-in simulation and found that they require an extremely low dust-to-metal ratio \citep[0.08, as oppose to the canonical value of 0.4 in the local Universe;][]{Dwek:1998} and high dust temperature ($91\pm23$\,K) in order to reproduce the SED of the $z\sim8.38$ dusty galaxy detected by \cite{Laporte:2017a}. Their results suggest that galaxies at such high redshifts are likely dust poor and have very high dust temperature. Population-wise, \cite{Cen:2014} have run dust radiative transfer calculations on a sample of 198 galaxies in a cosmological zoom-in simulation at $z\sim7$ and predicted dust luminosities, SEDs, and IRLF at $z\sim7$. They found that 60--90\% of the starlight is re-emitted in the IR, with the peak wavelength of dust SED around 45--60\,$\mu$m.

Note that large-volume cosmological simulations usually have mass resolution $\sim10^6\,\Msun$ and spatial resolution $\sim1$\,kpc. Even the zoom-in simulations discussed above are only able to resolve down to $\sim30$\,pc. It should be noted that dust geometry (clumpiness and covering fraction) and relative distribution between dust and stars have dramatic effects on the effective dust attenuation law, even if dust properties are constant \citep[e.g.][]{Seon:2016,Narayanan:2018a}. Moderate- and low-resolution simulations sometimes adopt sub-resolution models to account for dust distribution on unresolved scales \citep[e.g.][]{Jonsson:2010}, especially the heavy obscuration of young stars from their birth clouds \citep[e.g.][]{Charlot:2000}. These models introduce extra free parameters which the results can be sensitive to \citep[e.g.][]{Cen:2014,Liang:2018}.

In this work, we present a suite of high-resolution cosmological zoom-in simulations of $z\geq5$ galaxies as part of the Feedback in Realistic Environments (FIRE) project\footnote{https://fire.northwestern.edu} \citep{Hopkins:2014,Hopkins:2018b}. These simulations cover a broad range of halo mass up to $10^{12}\,\Msun$ at $z=5$--10. We adopt a mass resolution $\sim7000\,\Msun$ or better and the typical spatial resolution in dense gas is $\sim1$\,pc. They use the FIRE-2 models of the multi-phase interstellar medium (ISM), star formation, and stellar feedback, which explicitly resolve stars forming in birth clouds and feedback disrupting these clouds. Simulations run down to $z\sim0$ using these models have been shown to reproduce a variety of observables, including the properties of giant molecular clouds (GMCs) in the local Universe \citep[see][and references therein]{Hopkins:2018b}. In particular, the $z\geq5$ simulations are shown to produce reasonable stellar mass--halo mass relation, SFR--stellar mass relation, stellar mass functions, UVLFs, and cosmic SFRD that are broadly consistent with most up-to-date observational constraints \citep{Ma:2018a}.

By post-processing these simulations with Monte Carlo dust radiative transfer calculations (without the need of sub-resolution dust recipes), we will study dust attenuation and emission in $z\geq5$ galaxies that would be detectable in wide-field deep surveys in the rest-frame UV. Our work is built upon previous theoretical studies on dusty galaxies at $z\geq5$ by expanding the sample size, using more detailed simulations and post-processing methods, and broadening the scope. The paper is organized as follows. We briefly describe our simulation sample and the baryonic physics used in these simulations in Section \ref{sec:sim}. In Section \ref{sec:rt}, we describe the radiative transfer calculations. Section \ref{sec:result} mainly focuses on the UV and IR properties of dusty galaxies at $z\geq5$, where we investigate the IRX--$\buv$ relation in Section \ref{sec:irx}, the bolometric luminosity of dust emission and its correlation with star formation activities in Section \ref{sec:bolIR}, and dust SEDs and dust temperature in Section \ref{sec:sed}. Section \ref{sec:lf} focuses on predicting galaxy UV and bolometric IR LFs and cosmic SFRD at $z=5$--10. We discuss the strategies for probing dusty $z\geq5$ galaxies and the limitations of this work in Section \ref{sec:discussion}. We conclude in Section \ref{sec:conclusion}.

We adopt a standard flat {\LCDM} cosmology with {\it Planck} 2015 cosmological parameters $H_0=68 {\rm\,km\,s^{-1}\,Mpc^{-1}}$, $\Omega_{\Lambda}=0.69$, $\Omega_{m}=1-\Omega_{\Lambda}=0.31$, $\Omega_b=0.048$, $\sigma_8=0.82$, and $n=0.97$ \citep{Planck-Collaboration:2016a}. We use a \citet{Kroupa:2002} initial mass function (IMF) from 0.1--$100\,\Msun$, with IMF slopes of $-1.30$ from 0.1--$0.5\,\Msun$ and $-2.35$ from 0.5--$100\,\Msun$. All magnitudes are in the AB system \citep{Oke:1983}.

\section{Methods}
\label{sec:method}

\begin{table*}
\caption{Simulation details.}
\begin{threeparttable}
\begin{tabular}{cccccccccccccc}
\hline
Name & $z_{\rm final}$ & $\Mhalo$ & $\mb$ & $\mDM$ & $\eg$ & $\eDM$ & $\Ms$ & $\Mdust$ & $\Lbol$ & $\LIR$ & $\rm SFR_{10}$ & $\rm SFR_{100}$ & $\Teff$ \\
 & & [$\Msun$] & [$\Msun$] & [$\Msun$] & [pc] & [pc] & [$\Msun$] & [$\Msun$] & [$\Lsun$] & [$\Lsun$] & [$\Msun\,\yr^{-1}$] & [$\Msun\,\yr^{-1}$] & [K] \\
\hline
z5m12b & 5 & 8.73e11 & 7126.5 & 3.9e4 & 0.42 & 42 & 2.55e10 & 1.31e8 & 1.57e12 & 1.01e12 & 170.6 & 70.75 & 34.72 \\
z5m12c & 5 & 7.91e11 & 7126.5 & 3.9e4 & 0.42 & 42 & 1.83e10 & 1.31e8 & 9.91e11 & 6.14e11 & 81.73 & 52.68 & 31.87 \\
z5m12d & 5 & 5.73e11 & 7126.5 & 3.9e4 & 0.42 & 42 & 1.20e10 & 7.28e7 & 3.37e11 & 8.31e10 & 8.79 & 46.57 & 25.06 \\
z5m12e & 5 & 5.04e11 & 7126.5 & 3.9e4 & 0.42 & 42 & 1.35e10 & 6.30e7 & 1.04e12 & 5.99e11 & 118.3 & 50.22 & 36.36 \\
z5m12a & 5 & 4.51e11 & 7126.5 & 3.9e4 & 0.42 & 42 & 5.36e9 & 3.25e7 & 2.17e11 & 9.33e10 & 18.24 & 9.53 & 29.26 \\
z5m11f & 5 & 3.15e11 & 7126.5 & 3.9e4 & 0.42 & 42 & 4.68e9 & 2.53e7 & 6.90e11 & 3.33e11 & 88.78 & 22.17 & 37.38 \\
z5m11e & 5 & 2.47e11 & 7126.5 & 3.9e4 & 0.42 & 42 & 2.53e9 & 1.97e7 & 6.25e10 & 1.02e10 & 1.76 & 10.21 & 21.86 \\
z5m11g & 5 & 1.98e11 & 7126.5 & 3.9e4 & 0.42 & 42 & 1.86e9 & 1.40e7 & 8.56e10 & 3.67e10 & 7.15 & 4.71 & 28.52 \\
z5m11d & 5 & 1.35e11 & 7126.5 & 3.9e4 & 0.42 & 42 & 1.62e9 & 7.54e6 & 3.66e10 & 8.14e9 & 1.81 & 3.31 & 24.67 \\
z5m11h & 5 & 1.01e11 & 7126.5 & 3.9e4 & 0.42 & 42 & 1.64e9 & 7.78e6 & 5.77e10 & 9.45e9 & 3.31 & 5.49 & 25.09 \\
z5m11c & 5 & 7.62e10 & 7126.5 & 3.9e4 & 0.42 & 42 & 7.52e8 & 2.57e6 & 1.42e10 & 1.28e9 & 0.55 & 1.18 & 21.64 \\
z5m11i & 5 & 5.47e10 & 7126.5 & 3.9e4 & 0.42 & 42 & 3.39e8 & 1.19e6 & 5.39e9 & 1.64e8 & 0.040 & 0.18 & 17.52 \\
z5m11b & 5 & 4.02e10 & 954.4 & 5.2e3 & 0.28 & 21 & 1.67e8 & 5.55e5 & 1.14e10 & 2.99e9 & 1.32 & 0.22 & 31.86 \\
z5m11a & 5 & 4.16e10 & 954.4 & 5.2e3 & 0.28 & 21 & 1.22e8 & 4.71e5 & 4.20e9 & 2.29e8 & 0.30 & 0.28 & 21.35 \\
z5m10f & 5 & 3.30e10 & 954.4 & 5.2e3 & 0.28 & 21 & 1.56e8 & 7.47e5 & 2.92e9 & 2.14e7 & 0.012 & 0.69 & 13.57 \\
z5m10e & 5 & 2.57e10 & 954.4 & 5.2e3 & 0.28 & 21 & 3.93e7 & 3.20e5 & 3.41e9 & 2.67e8 & 0.30 & 0.19 & 23.29 \\
z5m10d & 5 & 1.87e10 & 954.4 & 5.2e3 & 0.28 & 21 & 4.81e7 & 2.52e5 & 1.77e9 & 3.07e7 & 0.049 & 0.22 & 16.95 \\
z5m10c & 5 & 1.34e10 & 954.4 & 5.2e3 & 0.28 & 21 & 5.58e7 & 2.34e5 & 5.14e9 & 1.68e8 & 0.35 & 0.34 & 22.78 \\
z5m10b & 5 & 1.25e10 & 954.4 & 5.2e3 & 0.28 & 21 & 3.42e7 & 8.88e4 & 4.61e9 & 6.83e8 & 0.55 & 0.066 & 33.62 \\
z5m10a & 5 & 6.86e9 & 954.4 & 5.2e3 & 0.28 & 21 & 1.58e7 & 5.07e4 & 2.74e8 & 4.78e6 & 0.013 & 0.018 & 16.27 \\
z5m09b & 5 & 3.88e9 & 119.3 & 650.0 & 0.14 & 10 & 2.79e6 & 1.34e4 & 3.36e7 & 1.98e4 & 2.08e-4 & 0.002 & 8.33 \\
z5m09a & 5 & 2.36e9 & 119.3 & 650.0 & 0.14 & 10 & 1.64e6 & 1.11e4 & 3.56e7 & 1.41e4 & 2.12e-4 & 0.008 & 8.09 \\
z7m12a & 7 & 8.91e11 & 7126.5 & 3.9e4 & 0.42 & 42 & 1.66e10 & 8.63e7 & 1.60e12 & 1.14e12 & 161.0 & 83.04 & 38.78 \\
z7m12b & 7 & 6.40e11 & 7126.5 & 3.9e4 & 0.42 & 42 & 1.44e10 & 5.59e7 & 1.06e12 & 6.79e11 & 95.87 & 56.05 & 38.55 \\
z7m12c & 7 & 4.71e11 & 7126.5 & 3.9e4 & 0.42 & 42 & 1.16e10 & 5.65e7 & 1.16e12 & 5.73e11 & 114.1 & 71.01 & 36.00 \\
z7m11a & 7 & 3.32e11 & 7126.5 & 3.9e4 & 0.42 & 42 & 7.17e9 & 3.50e7 & 3.43e11 & 9.95e10 & 16.62 & 41.70 & 29.50 \\
z7m11b & 7 & 2.48e11 & 7126.5 & 3.9e4 & 0.42 & 42 & 2.00e9 & 1.24e7 & 2.75e11 & 1.25e11 & 32.57 & 12.03 & 35.70 \\
z7m11c & 7 & 1.63e11 & 7126.5 & 3.9e4 & 0.42 & 42 & 1.81e9 & 1.03e7 & 6.58e10 & 1.08e10 & 4.58 & 4.09 & 24.52 \\
z9m12a & 9 & 4.20e11 & 7126.5 & 3.9e4 & 0.42 & 42 & 1.24e10 & 5.67e7 & 1.35e12 & 1.13e12 & 141.4 & 65.30 & 42.08 \\
z9m11a & 9 & 2.88e11 & 7126.5 & 3.9e4 & 0.42 & 42 & 3.46e9 & 1.62e7 & 6.23e11 & 2.35e11 & 59.73 & 26.85 & 37.91 \\
z9m11b & 9 & 2.23e11 & 7126.5 & 3.9e4 & 0.42 & 42 & 3.49e9 & 1.35e7 & 2.98e11 & 1.02e11 & 23.60 & 21.79 & 34.27 \\
z9m11c & 9 & 1.76e11 & 7126.5 & 3.9e4 & 0.42 & 42 & 2.41e9 & 9.81e6 & 1.66e11 & 7.22e10 & 14.13 & 13.45 & 34.26 \\
z9m11d & 9 & 1.28e11 & 7126.5 & 3.9e4 & 0.42 & 42 & 1.46e9 & 5.00e6 & 4.45e10 & 6.76e9 & 1.94 & 3.38 & 25.55 \\
z9m11e & 9 & 1.16e11 & 7126.5 & 3.9e4 & 0.42 & 42 & 1.49e9 & 7.58e6 & 8.73e10 & 2.09e10 & 4.12 & 11.50 & 28.82 \\
\hline
\end{tabular}
\begin{tablenotes}
\item Parameters describing the initial conditions and final galaxy properties of our simulations:
\item (1) $z_{\rm final}$: The redshift which the zoom-in region is selected at and the simulation is run to.
\item (2) $\Mhalo$: Halo mass of the central halo at $z_{\rm final}$.
\item (3) $\mb$ and $\mDM$: Initial baryonic and DM particle mass in the high-resolution region. The masses of DM particles are fixed throughout the simulation. The masses of baryonic (gas and stars) particles are allowed to vary within a factor of two owing to mass loss and mass return due to stellar evolution.
\item (4) $\eg$ and $\eDM$: Plummer-equivalent force softening lengths for gas and DM particles, in comoving units above $z=9$ and physical units thereafter. Force softening for gas is adaptive ($\eg$ is the minimum softening length). Force softening length for star particles is $\es=5\eg$.
\item (5) $\Ms$ and $\Mdust$: Total stellar and dust mass within the virial radius, assuming a constant dust-to-metal ratio of $\fdust=0.4$ in gas below $10^6$\,K and no dust in gas hotter than $10^6$\,K.
\item (6) $\Lbol$ and $\LIR$: Bolometric and dust IR luminosity integrated from 0.08--1000\,$\mu$m, accounting for all light coming out from the virial radius.
\item (7) $\rm SFR_{10}$ and $\rm SFR_{100}$: Star formation rate averaged over the past 10 and 100\,Myr, respectively, measured within the viral radius.
\item (8) $\Teff$: Dust effective temperature $\Teff=(\int\rho T_{\rm eq}^{4+\beta}\dd V/\int\rho\dd V)^{1/(4+\beta)}\sim (\LIR/\Mdust)^{1/(4+\beta)}$, where $\beta=2$ is the dust emissivity spectral index and $T_{\rm eq}$ is the dust temperature assuming local thermal equilibrium (LTE).
\end{tablenotes}
\end{threeparttable}
\label{tbl:sim}
\end{table*}

\subsection{The simulations}
\label{sec:sim}
This work uses a suite of 34 high-resolution cosmological zoom-in simulations at $z\geq5$. The zoom-in regions are centered around halos randomly selected at desired mass and redshift from a set of dark matter (DM)-only cosmological boxes with periodic boundary conditions. The initial conditions are generated at $z=99$ following the method in \cite{Onorbe:2014} using the {\sc music} code \citep{Hahn:2011}, which uses the well-developed multi-scale cosmological zoom-in technique \citep[e.g.][]{Katz:1993,Bertschinger:2001}. We ensure zero contamination from low-resolution particles within $2\Rvir$ of the central halo, and less than 1\% contamination in $3\Rvir$. 22 zoom-in regions are selected from a $(30\,\hinv\Mpc)^3$ box run to $z=5$ around halos in $\Mhalo\sim10^{9.5}$--$10^{12}\,\Msun$, among which 15 are first presented in \cite{Ma:2018a} and 7 more are added to improve the statistics at the high-mass end. Another 6 zoom-in regions are selected from a $(120\,\hinv\,\Mpc)^3$ box run to $z=7$ and the rest 6 from an independent box with the same size run to $z=9$. They are centered on relatively more massive halos from $\Mhalo\sim10^{11}$--$10^{12}\,\Msun$ at $z=7$ and $z=9$, respectively.

The initial mass for baryonic particles (gas and stars) ranges from $\mb=100$--$7000\,\Msun$, and high-resolution DM particles from $\mDM=650$--$4\times10^4\,\Msun$, increasing with the mass of the central halo. Force softening for gas particles is adaptive, with a minimum Plummer-equivalent force softening length $\eg=0.14$--$0.42\,\pc$. Force softening lengths for star particles and high-resolution DM particles are fixed at $\es=5\eg=0.7$--2.1\,pc and $\eDM=10$--42\,pc, respectively. The softening lengths are in comoving units at $z>9$ and in physical units thereafter. In Table \ref{tbl:sim}, we provide the final redshift, mass resolution, force softening lengths, final halo mass, stellar mass, and selected galaxy properties of the central halo for all 34 zoom-in simulations. We explicitly check and confirm that there is no systematic difference between galaxies of similar masses but simulated at different resolution in all of our results in this paper (examples shown in Appendix \ref{app:conv}).

All simulation are run using an identical version of the code {\sc gizmo}\footnote{http://www.tapir.caltech.edu/{\textasciitilde}phopkins/Site/GIZMO.html} \citep{Hopkins:2015} in the meshless finite-mass (MFM) mode with the FIRE-2 models of the multi-phase ISM, star formation, and stellar feedback \citep{Hopkins:2018b}, which we briefly summarize here. Gas follows an ionized+atomic+molecular cooling curve in 10--$10^{10}$\,K, including metallicity-dependent fine-structure and molecular cooling at low temperatures and high-temperature metal-line cooling for 11 separately tracked species (H, He, C, N, O, Ne, Mg, Si, S, Ca, and Fe). At each timestep, the ionization states and cooling rates for H and He are computed following \cite{Katz:1996} and cooling rates from heavier elements are calculated from a compilation of {\sc cloudy} runs \citep{Ferland:2013}, applying a uniform, redshift-dependent ionizing background from \cite{Faucher-Giguere:2009} and an approximate model for H\,{\sc ii} regions generated by local sources. Gas self-shielding is accounted for with a local Jeans-length approximation. 

Star formation is only allowed in dense, molecular, and locally self-gravitating regions with hydrogen number density above $\nc=1000\,\cm^{-3}$ \citep{Hopkins:2013b}. Each star particles is treated as a stellar population with known mass, age, and metallicity assuming a \cite{Kroupa:2002} initial mass function (IMF) from 0.1--$100\,\Msun$. The simulations account for the following feedback mechanisms: (1) local and long-range radiation pressure, (2) photoionization and photoelectric heating, and (3) energy, momentum, mass, and metal injection from supernovae (SNe) and stellar winds. The luminosity, mass loss rates, and Type-II SNe rates of each star particle are obtained from {\sc starburst99} \citep{Leitherer:1999}, and Type-Ia SNe rates following \cite{Mannucci:2006}. Metal yields from Type-II and Ia SNe and AGB winds are taken from \cite{Nomoto:2006}, \cite{Iwamoto:1999}, and \cite{Izzard:2004}, respectively. All simulations\footnote{The simulations presented in \cite{Ma:2018a} are rerun from the same initial conditions with sub-resolution metal diffusion.} are run with a sub-resolution turbulent metal diffusion algorithm described in \cite{Su:2017} and \cite{Escala:2018}. We do not account for primordial chemistry nor Pop III star formation, but assume an initial metallicity of $Z=10^{-4}\,\Zsun$.

\begin{figure}
\centering
\includegraphics[width=\linewidth]{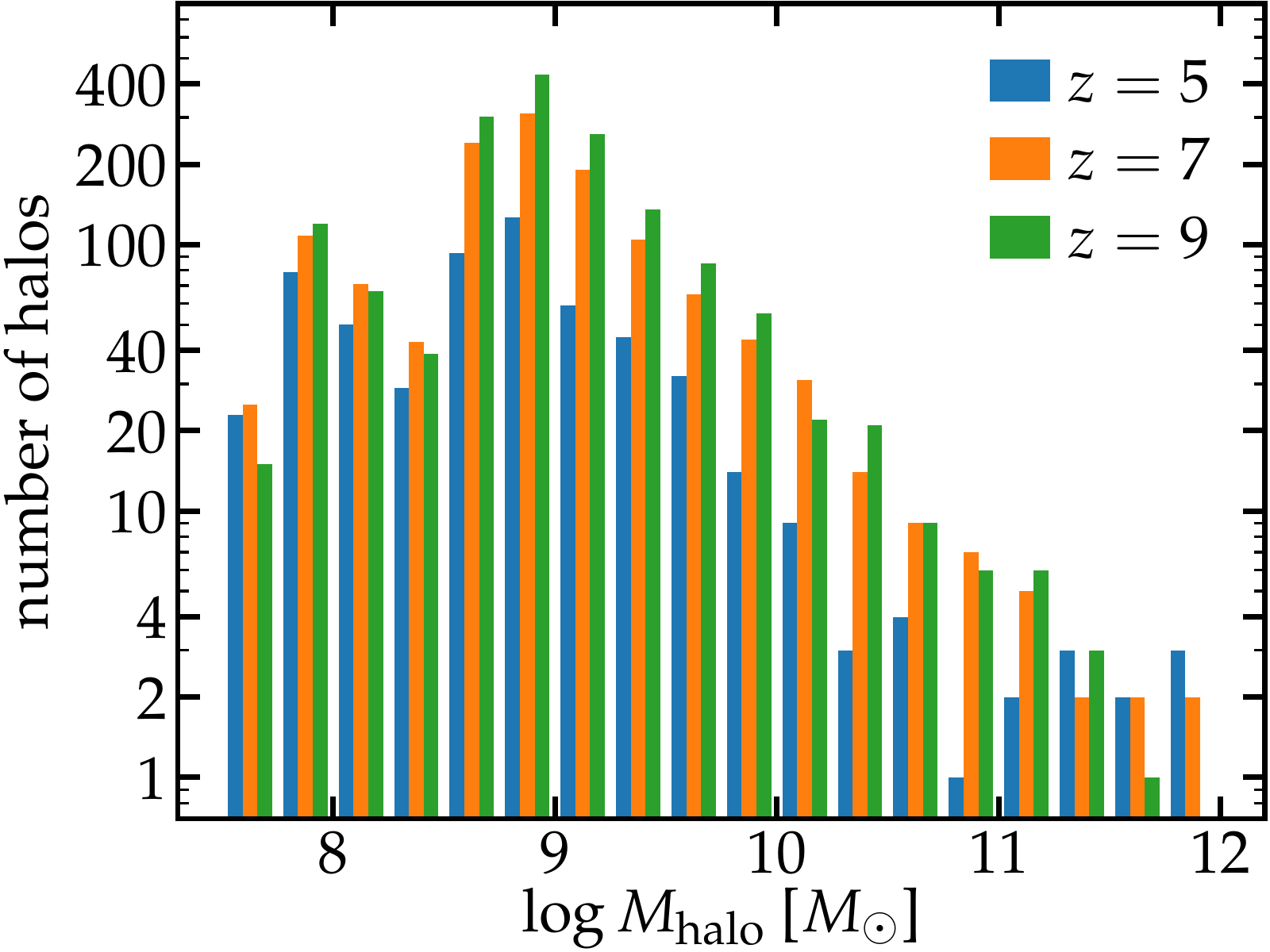}
\caption{Number of sufficiently resolved halos in every $\Delta\log\Mhalo=0.25$ dex in our simulation sample at $z=5$, 7, and 9. These halos will be used to derive rest-frame UVLFs in Section \ref{sec:uvlf}. Dust radiative transfer calculations are conducted in all halos more massive than $10^{10}\,\Msun$ and a small number of halos above $10^{9.5}\,\Msun$.}
\label{fig:nhalo}
\end{figure}

We use the Amiga's halo finder \citep[{\sc ahf};][]{Knollmann:2009} to identify halos and galaxies in the snapshots, applying the redshift-dependent virial parameter from \cite{Bryan:1998}. There are more than one halo in each zoom-in region. In this paper, we restrict our analysis to halos that contain more than $10^4$ particles and have zero contamination from low-resolution particles within $\Rvir$ to ensure good resolution. We also exclude subhalos from our study. In Fig. \ref{fig:nhalo}, we show the number of halos selected based on the criteria above in every 0.25\,dex from $\log\Mhalo=7.5$--12 at $z=5$, 7, and 9. These halos will be used to derive the rest-frame UVLFs in Section \ref{sec:uvlf}. The dust radiative transfer calculations described below are only conducted for all halos more massive than $\Mhalo=10^{10}\,\Msun$ and central halos above $10^{9.5}\,\Msun$, as dust is negligible in halos of lower masses. We also include all snapshots in our analysis and treat them as independent galaxies, which we refer as `galaxy snapshots' below ($\sim20\,\Myr$ between snapshots), to account for short-time-scale variabilities of galaxy properties due to bursty star formation in our simulations \citep{Ma:2018a}.

\subsection{Basic properties of the simulated galaxies}
\label{sec:property}
\begin{figure}
\centering
\includegraphics[width=\linewidth]{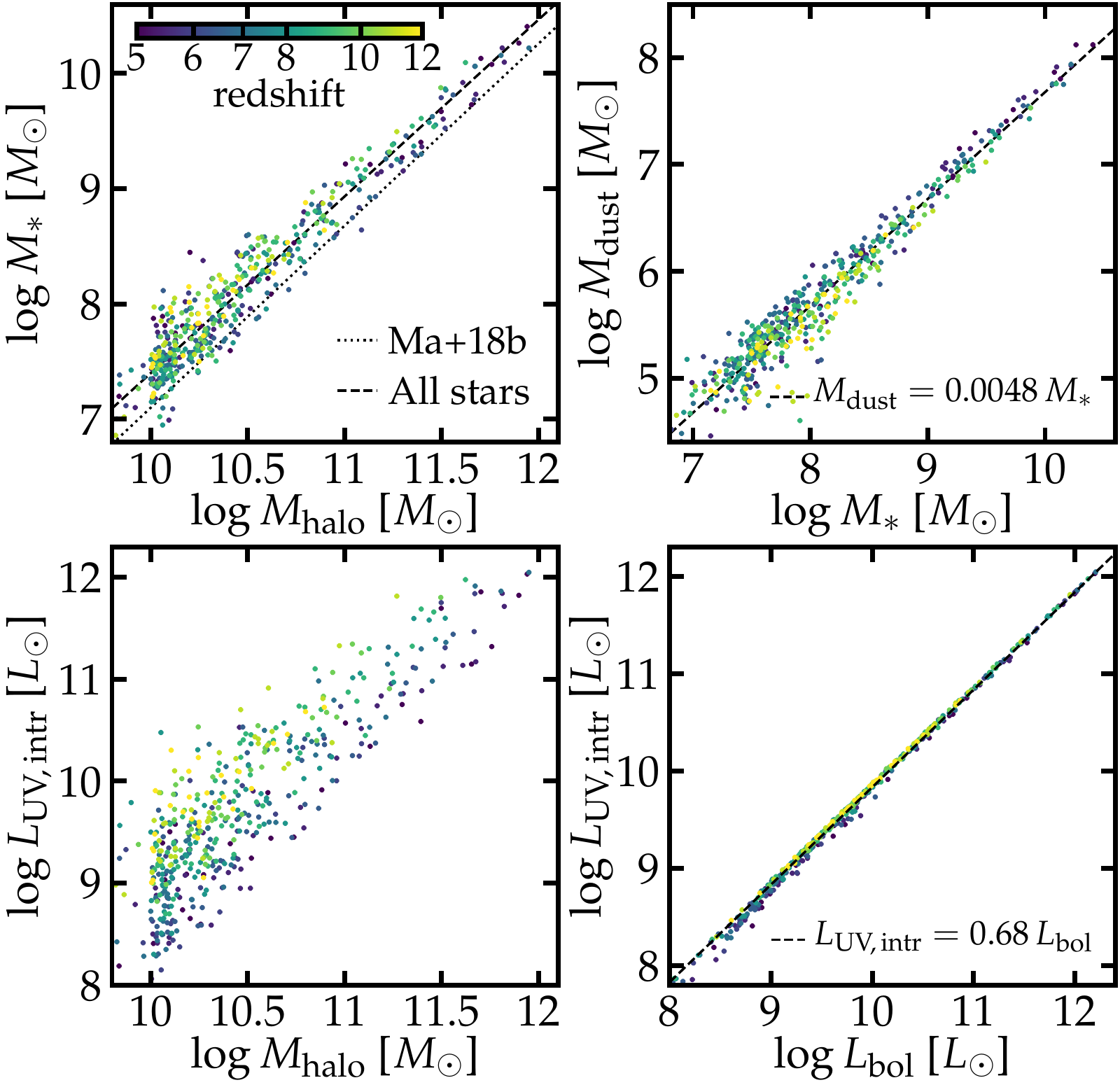}
\caption{Scaling relations of simulated galaxies. {\em Top left}: The stellar mass--halo mass relation. Here we use the total stellar mass in $\Rvir$, so the best-fit linear region (dashed line) lies above the one from \citet[][dotted line]{Ma:2018a}, where $\Ms$ is measured in a much smaller radius to exclude satellites and diffuse stars. {\em Top right}: The dust mass--stellar mass relation. There is a linear correlation $\Mdust=0.0048\,\Ms$, as dust is produced by stars following our assumptions. {\em Bottom left}: The {\em intrinsic} (unobscured) UV luminosity--halo mass relation. At a given halo mass, $\LUVi$ increases with redshift by an order of magnitude from $z=5$ to 12, as galaxies at higher redshift tend to have higher SFRs. {\em Bottom right}: Intrinsic UV luminosity--bolometric luminosity relation. $\LUVi$ is a good proxy of $\Lbol$ with little scatter ($\LUVi=0.68\,\Lbol$), as young stars dominate both the UV luminosity and the total luminosity. We will use these relations to interpret the results in the rest of this paper.}
\label{fig:smhm}
\end{figure}

In Fig. \ref{fig:smhm}, we present scaling relations for our simulated galaxies at integer redshifts from $z=5$--12. Each point represents one galaxy snapshot in our sample, color-coded by its redshift. The top-left panel shows the stellar mass--halo mass relation, where we use the total stellar mass within $\Rvir$. The dashed line shows the best-fit linear relation $\log\Ms=1.53\,(\log\Mhalo-10)+7.40$. The dotted line shows the linear fit from \cite{Ma:2018a}. Note that they measure stellar mass within $\Rmax/3$ to exclude satellite galaxies and diffuse stars, where $\Rmax$ is the halo maximum velocity radius and $\Rmax/3$ is roughly comparable to $0.2\,\Rvir$ in these halos. There is thus a 0.2--0.3\,dex difference between the two relations. The redshift evolution of the $\Ms$--$\Mhalo$ relation is not significant (by less than 0.1\,dex from $z=5$ to $z=12$, still within the scatter of the sample).

The top-right panel shows the relation between dust mass and stellar mass, both measured within $\Rvir$. Here we assume a constant dust-to-metal ratio of 0.4 \citep{Dwek:1998} in gas below $10^6$\,K and no dust in hotter gas. Unsurprisingly, $\Mdust$ is proportional to $\Ms$ (dust mass equals to 0.48\% of the stellar mass), as all the dust is produced in stellar evolution processes (SNe and AGB stars) by assumption. The bottom-left panel presents the relation between the {\em intrinsic} (unobscured) rest-frame UV luminosity ($\LUVi\equiv\lambda L_{\lambda}$ at 1500\,\AA, measured within $\Rvir$) and halo mass. At a given $\Mhalo$ (and $\Ms$), $\LUVi$ increases by an order of magnitude from $z=5$ to $z=12$, because galaxies at higher redshifts have higher SFRs \citep[see also][]{Ma:2018a}. The bottom-right panel shows the relation between $\LUVi$ and bolometric luminosity $\Lbol$ of stellar continuum. We find a universal relation $\LUVi=0.68\,\Lbol$ for our simulated sample with no discernible scatter, because young (UV-bright) stars also dominate the total luminosity. We will use these scaling relations to interpret our results in the rest of this paper.

\subsection{Dust radiative transfer}
\label{sec:rt}
We post-process our simulations with the public three-dimensional Monte Carlo dust radiative transfer code {\sc skirt}\footnote{http://www.skirt.ugent.be/root/index.html} \citep{Baes:2011,Camps:2015} to calculate galaxy continuous spectral energy distributions (SEDs) on a 90-point wavelength grid equally spaced in logarithmic scale from 0.08--1000\,$\mu$m. For each halo, we include all gas and star particles out to $\Rvir$ in our calculations and compute galaxy SEDs and mock images at each wavelength along five random lines of sight.

We do not explicitly model dust formation, growth, and destruction in our simulations, but simply assume a constant dust-to-metal ratio ($\Mdust=\fdust\,M_{\rm metal}$) in gas below $10^6$\,K and no dust in hotter gas. The dust grid is reconstructed from gas particles using the built-in octree grid in {\sc skirt} \citep{Saftly:2013,Saftly:2014}, where we include all particles in a cubic domain with a side length of $2\Rvir$ and adaptively refines the high-density region until the following criteria are met: (1) the dust mass in a cell does not exceed $10^{-6}$ of the total dust mass in the domain and (2) the $15^{\rm th}$ refinement level has reached (i.e. the cell size is $2^{-15}$ of the domain size). The minimum cell width is less than 3\,pc even for the most massive galaxy in our sample. We use $10^6$ photon packets at each of the 90 wavelengths. These choices ensure excellent convergence at a reasonable computational cost. We refer to Appendix \ref{app:conv} for details about the convergence tests.

\begin{figure}
\centering
\includegraphics[width=\linewidth]{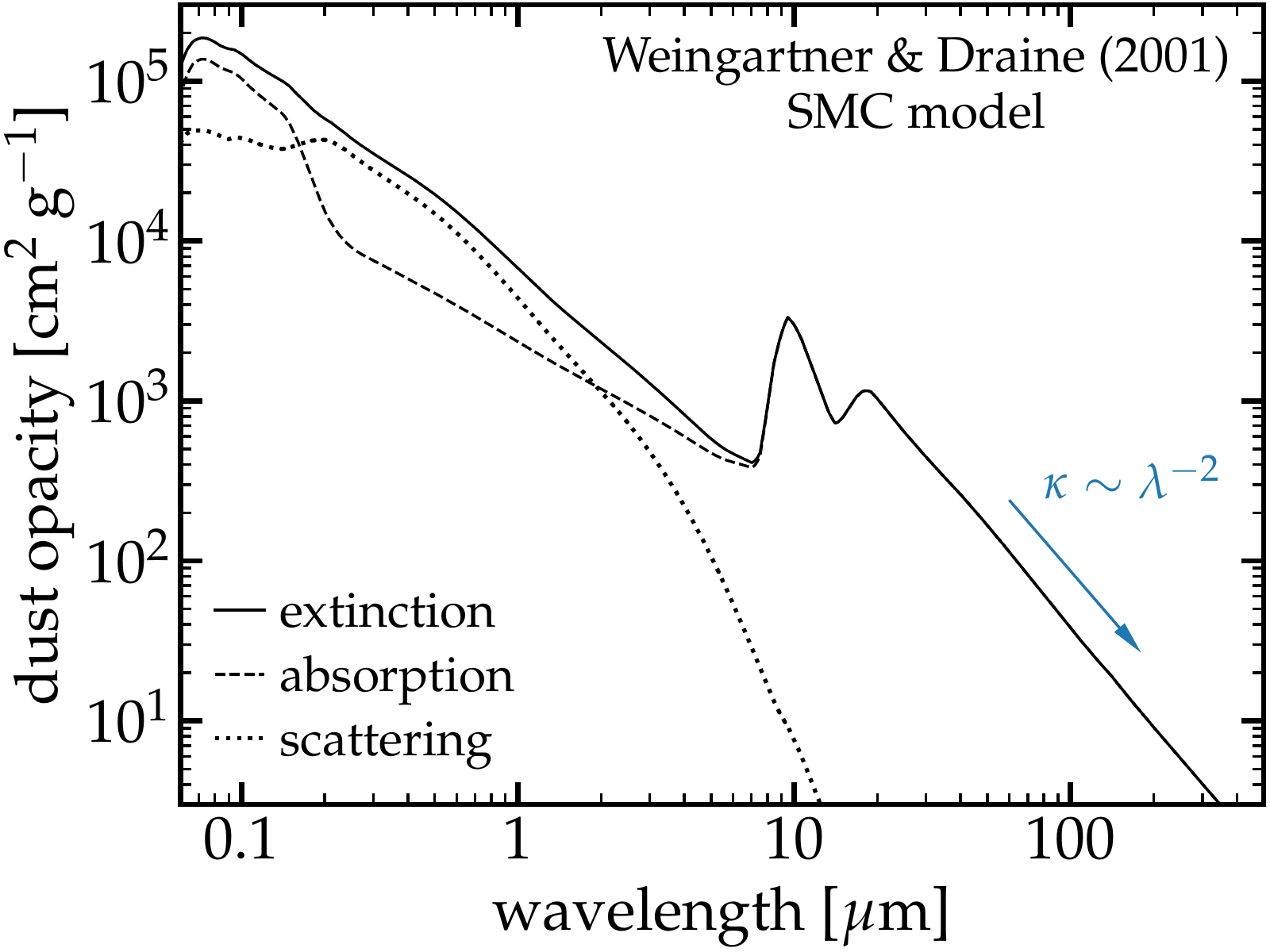}
\caption{Dust opacity for the SMC grain size distribution model in \citet{Weingartner:2001}. Dust absorption and scattering are both important in the UV and optical. In the mid- and far-IR ($\lambda>30\,\mu$m), dust opacity scales with wavelength (frequency) as $\kappa\propto\lambda^{-2}$ ($\nu^2$).}
\label{fig:opacity}
\end{figure}

\begin{figure*}
\centering
\includegraphics[width=\linewidth]{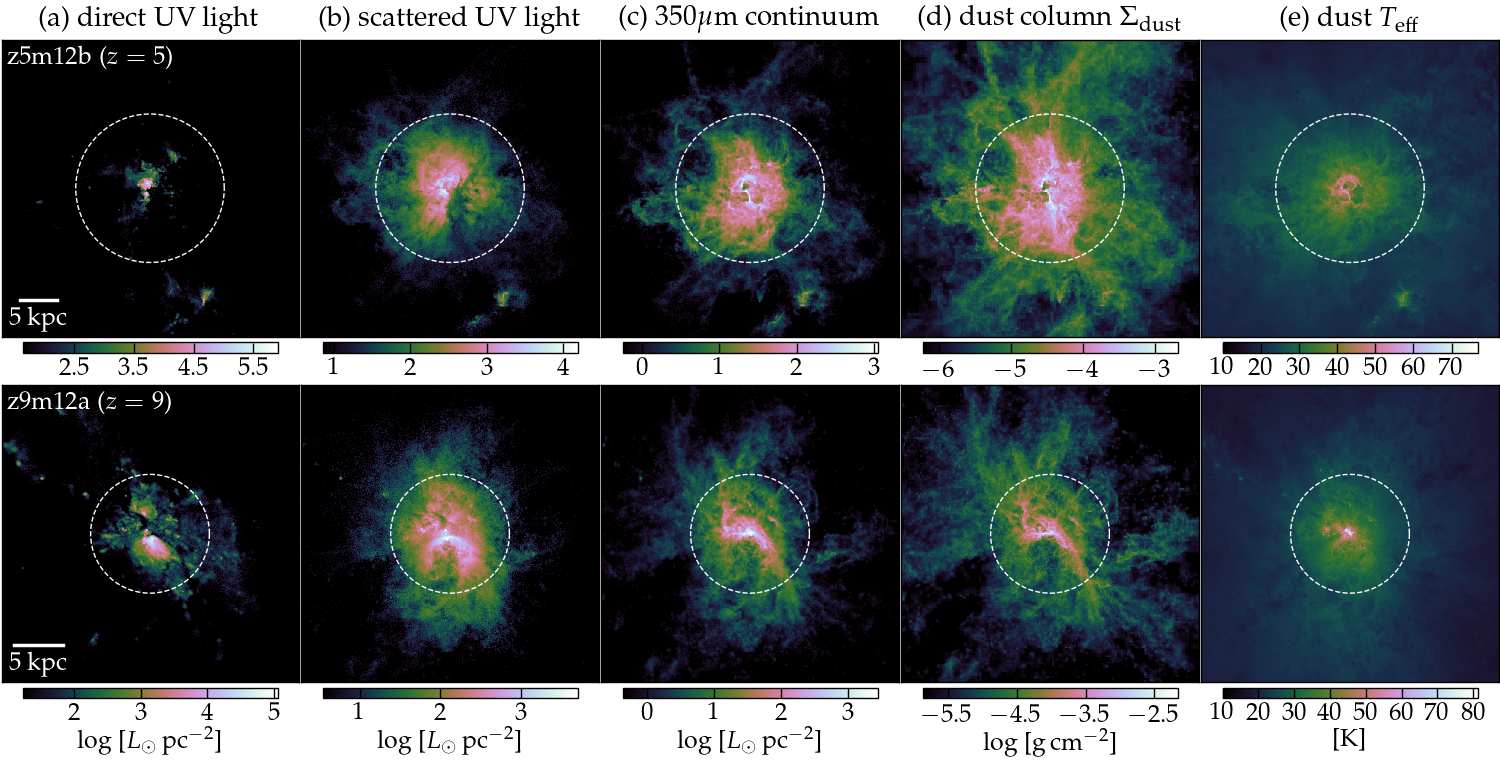}
\caption{Example images of two galaxies in our sample, z5m12b at $z=5$ (top) and z9m12a at $z=9$ (bottom). From left to right: (a) rest-frame UV continuum (1500\,\AA) directly transmitted from stars, (b) UV continuum from dust scattering, (c) rest-frame 350$\mu$m dust continuum, (d) dust column density, and (e) dust effective temperature. The white dashed circles show the $\Rmax/3$ radius. Dust distribution is patchy and extended to large radii. Scattered UV light contributes $\sim1/3$ of the post-extinction UV flux but is distributed over larger spatial scales and at lower surface brightness than the UV light direct from stars.}
\label{fig:image}
\end{figure*}

We adopt the Small Magellanic Cloud (SMC)-type dust grain size distribution from \citet[][table 3 therein]{Weingartner:2001}. In this model, silicate dust dominates the dust opacity and carbonaceous dust has a small contribution, as suggested for high-redshift systems \citep[e.g.][]{Dwek:2014}, but there is no polycyclic aromatic hydrocarbon (PAH) included. We also compare Milky Way (MW)-type dust in Appendix \ref{app:conv}. In Fig. \ref{fig:opacity}, we show the dust opacity from UV to far-IR. Dust absorption and scattering are both important in the UV and optical. At long wavelengths ($\lambda>30\,\mu$m), dust absorption dominates the opacity, which scales with wavelength (frequency) roughly as $\kappa\propto\lambda^{-\beta}$ ($\nu^{\beta}$) with dust emissivity spectral index $\beta=2$. Our fiducial dust-to-metal ratio is $\fdust=0.4$ \citep{Dwek:1998}, which gives a gas opacity\footnote{By definition, the relation between gas opacity due to dust extinction and dust grain opacity is simply $\kappa_{\rm gas}=\kappa_{\rm dust}\,\rho_{\rm dust}/\rho_{\rm gas}$.} at 1500\,\AA
\be
\label{eqn:opacity}
     \kappa_{1500\,{\text\AA}}=0.73\times10^3\,\cm^2\,\g^{-1} \left( \frac{\fdust}{0.4} \right) \left( \frac{Z_{\rm gas}}{\Zsun} \right),
\ee
assuming a solar metallicity $\Zsun=0.02$. In this work, we will also explore $\fdust=0.2$--0.8. We note that it is the absorption coefficient $\alpha \equiv \kappa_{\rm dust}\,\rho_{\rm dust} = \kappa_{\rm gas}\,\rho_{\rm gas}= \kappa_{\rm dust}\,\fdust\,Z_{\rm gas}\,\rho_{\rm gas}$ that enters the radiative transfer equation and sets the dust temperature and emissivity (via Kirchhoff's law). There is a degeneracy between dust opacity and dust-to-metal ratio in the form of $\kappa_{\rm dust}\,\fdust$ in these calculations. Our experiments with different $\fdust$ at fixed $\kappa_{\rm dust}$ should be understood as varying the normalization of the gas opacity.\footnote{Note that the SMC gas opacity at 1500\,{\AA} in \cite{Pei:1992} is approximately 154\,cm$^2$\,g$^{-1}$, a factor of two difference from Equation \ref{eqn:opacity} if using $0.1\,\Zsun$ for SMC metallicity (e.g. \citealt{Pei:1992} suggested that the B-band gas opacities in the MW and SMC follow roughly 10:1). A factor of a few variation is also seen between different lines of sight. Our experiments with $\fdust=0.2$--0.8 account for the uncertainties of both dust opacity and dust-to-metal ratio.} That said, if we vary $\kdust$ and fix $\fdust$, all radiative transfer results, including the intensity field and dust temperature, must be identical to our experiments here (varying $\fdust$ for fixed $\kdust$).

Photon packets are first launched from star particles and propagated in the domain until absorbed or escaped. The SEDs of star particles are calculated from the built-in {\sc starburst99} stellar population models in {\sc skirt} (nearly identical to those used in our simulations), which are compiled by \cite{Jonsson:2010} and include both stellar continuum and Balmer continuum from nebular emission. Dust temperature and emissivity are determined from the local intensity field assuming energy balance. Next, photon packets representing dust emission are launched and propagated in the domain. The local radiation field and dust temperature are then updated to account for dust self-absorption. This step is done iteratively until the dust SED converges within 1\%, when the calculation stops and a final solution is reached. In this work, we do not include heating from the CMB, but defer to a future study on its effects (see Section \ref{sec:limit} for more discussion).

We use non-local thermal equilibrium (NLTE) dust emission self-consistently calculated in {\sc skirt} \citep{Baes:2011,Camps:2015a}. This accounts for emission from small grains that are transiently heated by individual photons. Moreover, grains of different sizes are no longer at a single equilibrium temperature, but follow a temperature distribution. The NLTE dust emission only affects the dust SED at wavelengths shorter than rest-frame $30\,\mu$m. Nonetheless, {\sc skirt} still computes the equilibrium dust temperature for each cell ($T_{\rm eq}$) following 
\be
\label{eqn:lte}
    \int_0^{\infty} \kappa_{\nu}^{\rm abs} J_{\nu} \, \dd\nu = \int_0^{\infty} \kappa_{\nu}^{\rm abs} B_{\nu}(T_{\rm eq}) \,\dd\nu,
\ee
where $\kappa_{\nu}^{\rm abs}$ is the dust absorption opacity and $J_{\nu}$ is the local radiation intensity at frequency $\nu$. It is worth noting that the right-hand-side integral scales as $T_{\rm eq}^{4+\beta}$ given $\kappa_{\nu}^{\rm abs}\propto\nu^{\beta}$ at long wavelengths where $B_{\nu}(T_{\rm eq})$ dominates. We will still use $T_{\rm eq}$ to describe dust temperature in each cell, as the long-wavelength dust emission that we mainly focus on in this paper is not affected by NLTE effects.

We note that some previous works adopted the {\sc mappings iii} starburst SED models \citep{Groves:2008} for star particles younger than 10\,Myr to account for unresolved small-scale dust distribution \citep[e.g.][]{Camps:2016,Liang:2018}. These models describe the dynamic evolution of spherical {\HII} regions on spatial scales of $\sim5$--800\,pc around star clusters of mass $10^{3.5}$--$10^{7.5}\,\Msun$, which are born from at least 10 times more massive clouds. They also include dust extinction and emission from the photodissociation regions. Cosmological simulations at $10^5\,\Msun$ mass resolution or worse may use these models to account for sub-resolution dust distribution \citep[see the discussion in section 2 of][]{Jonsson:2010}. Our simulations, however, are able to resolve the mass and spatial scales at which the {\sc mappings iii} models describe. We thus do not use these models in our radiative transfer calculations, but take the dust distribution `as such' in the simulations as an alternative to the {\sc mappings iii} models \citep[see also][]{Behrens:2018}. All the results in this paper are fairly converged at the resolution of our simulations (we show examples in Appendix \ref{app:conv}).

\begin{figure}
\centering
\includegraphics[width=\linewidth]{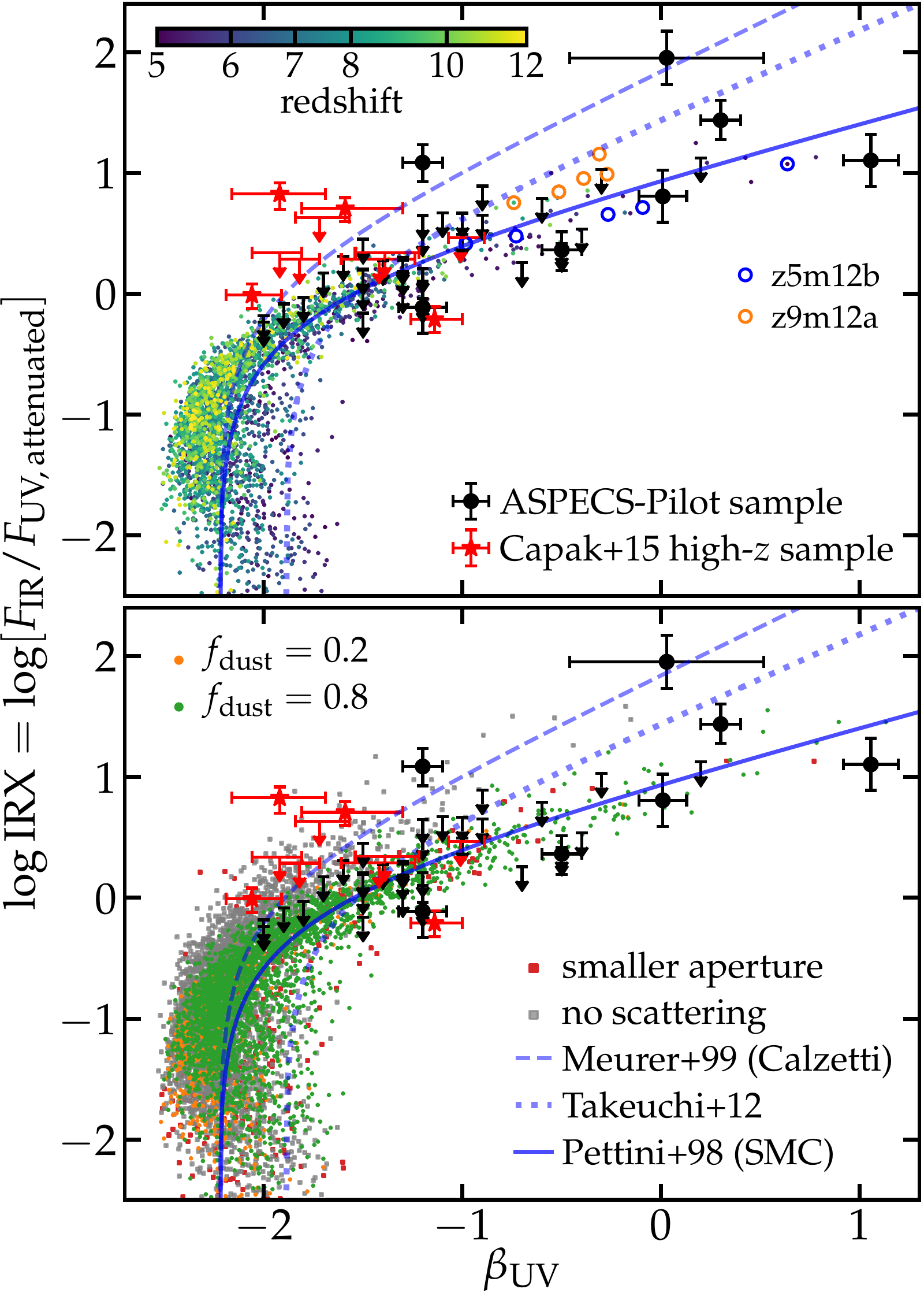}
\caption{{\em Top:} The IRX--$\buv$ relation for our simulations (using $\fdust=0.4$). Each point represents a galaxy snapshot along a random line of sight, color-coded by redshift. Galaxies at higher redshifts tend to move slightly toward bluer $\buv$ at fixed IRX due to their younger stellar populations. Points with errorbars show the observational datasets for high-redshift galaxies compiled in \citet[][consisting of two different samples]{Casey:2018a}. The lines show the empirical relations from the literature. The simulated sample forms a tight IRX--$\buv$ relation broadly agrees with observations and the SMC IRX--$\buv$ relation expected from a simple dust screen model. Our results suggest that patchy, complex dust distribution and non-trivial effect of dust scattering as shown in our simulations do not drive significant scatter in the IRX--$\buv$ relation. {\em Bottom:} The IRX--$\buv$ relation for different normalizations of the extinction curve (represented by varying $\fdust$; the color points). The grey squares show the results if dust scattering is ignored, which should be understood as changing the attenuation law. These results confirm that the IRX--$\buv$ relation is determined by the shape of the extinction curve but independent from its normalization. The red squares show the results if the IR and UV fluxes are measured in a smaller aperture ($\Rmax/3$ instead of $\Rvir$), which has little effect on the results.}
\label{fig:irx}
\end{figure}

\section{Results: dust attenuation and emission in high-redshift galaxies}
\label{sec:result}

\subsection{Example images}
\label{sec:example}
The built-in `peeling off' method (next-event estimator) and `smart detectors' \citep{Baes:2008} in {\sc skirt} can produce high signal-to-noise images with a relatively small number of photon packets. Thanks to its Monte Carlo nature, {\sc skirt} also allows us to separate light from different origins (e.g. from sources and dust, from direct transmission and scatter, etc.). In Fig. \ref{fig:image}, we present mock images for two example galaxies, z5m12b at $z=5$ (top) and z9m12a at $z=9$ (bottom), which are the most massive galaxy in our sample at each redshift (using $\fdust=0.4$). The color scale in columns (a)--(d) includes 95\% of the light/mass in the field of view. The white dashed circles show the $\Rmax/3$ radius.

Columns (a) and (b) show the rest-frame UV (1500\,\AA) images detected by a `smart camera', decomposed into (a) light transmitted directly from stars and (b) light scattered by dust at least once (i.e. adding (a) and (b) together gives the total UV flux as viewed by a regular camera). Most of the UV light is emitted from a compact region (less than 2\,kpc in projected radius) in the center of the galaxy. Some part of the galaxy is heavily obscured by optically-thick dust patches along the line of sight, while other part is almost transparent. Scattered UV light is more spatially extended and at lower surface brightness (note the different color scales), making it more difficult to detect observationally than light from direct transmission \citep[e.g.][]{Ma:2018}. For both galaxies, scattered light contributes 30--35\% of the total post-extinction UV flux.

Columns (c) and (d) show the rest-frame 350$\mu$m dust continuum image and projected dust column density, respectively. Dust distribution is clumpy and patchy and extends to a much larger spatial scale than stars, owing to feedback-driven outflows pushing gas and dust to large radii. For a dust opacity of 446.5 and 38.1\,cm$^2$\,g$^{-1}$ at 30 and $100\,\mu\rm m$, respectively (Fig. \ref{fig:opacity}), the mid- and far-IR is optically thin in most part of the galaxy except for the central dense region where the IR optical depth can reach order unity and dust self-absorption is thus important. Column (e) shows the dust effective temperature defined as $\Teff=(\int\rho T_{\rm eq}^{4+\beta}\dd l/\int\rho\dd l)^{1/(4+\beta)}$ with integration evaluated along the line of sight (see Section \ref{sec:sed} for motives for this definition). In the very central region that is close to the sources producing most of the UV light, dust can be heated to over 45\,K and a small fraction of dust (less than 0.1\% in mass) even reaches up to 100\,K. The diffuse dust out to 10\,kpc is also heated by diffuse starlight to 20--30\,K.

\begin{figure*}
\centering
\includegraphics[width=\linewidth]{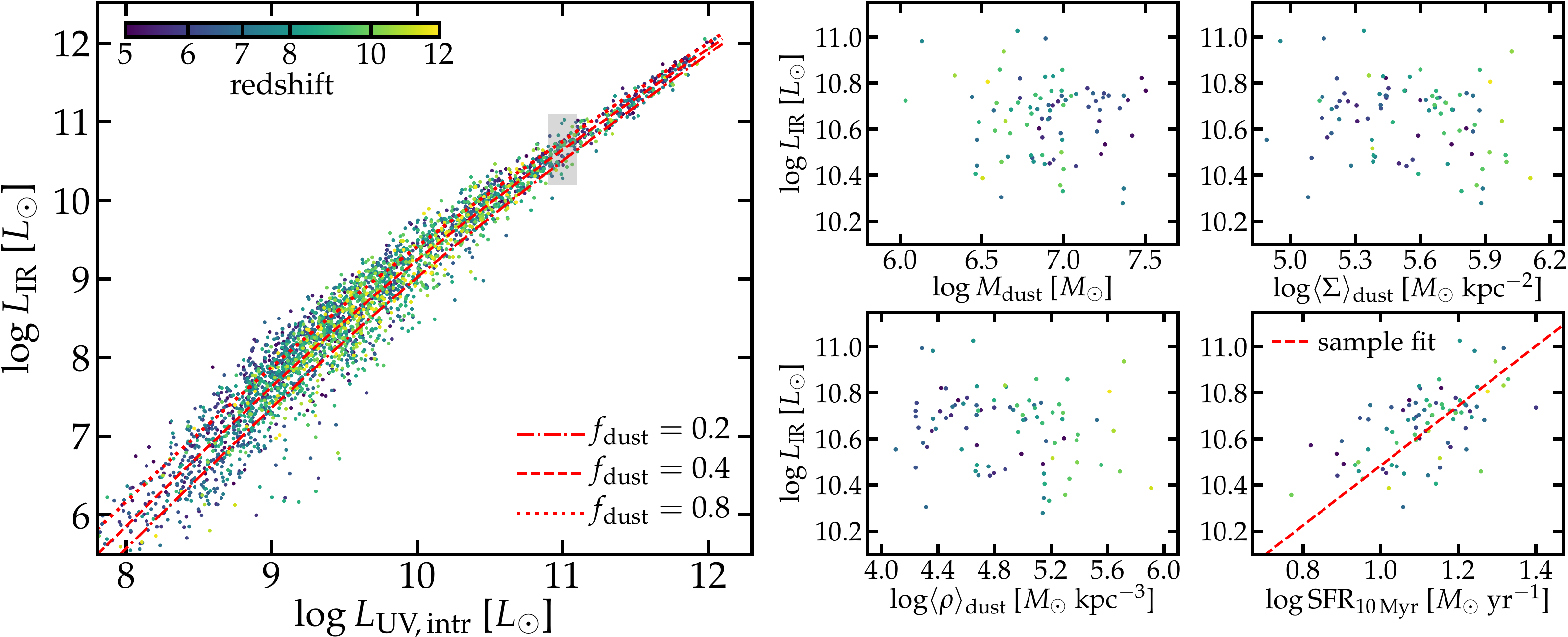}
\caption{{\em Left:} The relation between $\LIR$ and {\em intrinsic} $\LUVi$ for our sample. Each point represents one galaxy snapshot applying $\fdust=0.4$, color-coded by redshift. This relation does not depend on redshift and can be described by a broken power-law (Equation \ref{eqn:lirluv}, red dashed line), meaning that dust attenuation and emission become weaker for galaxies below $\LUVi\sim10^{10}\,\Lsun$ ($\rm M_{UV,\,intr}\sim-19$). The dependence of the $\LIR$--$\LUVi$ relation on $\fdust$ (normalization of the extinction curve) is illustrated by the red lines. At the bright end, $\LIR$ changes little with $\fdust$ (in the optically thick regime), whereas at the faint end where dust is optically thin, $\LIR$ is proportional to $\fdust$. {\em Right:} Secondary dependence of $\LIR$ on various properties. Each point is a galaxy snapshot within 0.1\,dex from $\LUVi=10^{11}\,\Lsun$ as marked by the grey shaded region in the left panel. At fixed $\LUVi$, $\LIR$ does not depend on total dust mass (top left), average dust column density (top right), and density (bottom left), suggesting that dust luminosity is primarily determined by dust covering fraction. The scatter in the $\LIR$--$\LUVi$ relation is driven by the amount of stars that are formed over the past 10\,Myr, indicating that young stars are more heavily obscured than relatively older stars.}
\label{fig:lirluv}
\end{figure*}

\subsection{The IRX--$\buv$ relation}
\label{sec:irx}
The relationship between the infrared excess, $\IRX=\FIR/\FUV$ ($\FUV$ is the attenuated UV flux here), and the rest-frame UV continuum slope, $\buv$ (to distinguish from the dust emissivity spectral index $\beta$ above), where $F_{\lambda}\sim\lambda^{\buv}$, is an empirical relation first established for local galaxies \citep{Meurer:1999} and being confirmed up to $z\sim2$ \citep[e.g.][]{Reddy:2012}. This relation is expected from a simple picture where an intrinsically blue source is obscured by a dust screen: as the amount of attenuation increases, the UV slope appears redder and the observed IR-to-UV flux ratio becomes larger. However, differential attenuation between young and old stars \citep[e.g.][]{Charlot:2000} or clumpiness of dust distribution \citep[e.g.][]{Seon:2016} can dramatically alter the effective attenuation law (even the dust composition is fixed), which may result in large variations in the IRX--$\buv$ relationship \citep[e.g.][]{Howell:2010,Casey:2014a,Narayanan:2018}.

At $z>3$, it is not yet clear whether the IRX--$\buv$ relation is still consistent with that for local galaxies \citep[e.g.][]{Koprowski:2018,McLure:2018}, where a Calzetti-like attenuation law should apply \citep{Calzetti:2000}, or follows a shallower relation that is more consistent with SMC-like attenuation law \citep[e.g.][]{Reddy:2018}, or there is no well-established IRX--$\buv$ relation due to large scatter \citep[e.g.][]{Capak:2015,Bouwens:2016,Barisic:2017}. It is also unclear whether such discrepancies in these observations are due to inconsistent measurements of the UV slope and IR luminosity \citep[e.g.][]{Gomez-Guijarro:2018}, selection biases in different samples, or intrinsic scatter driven by large variations of dust properties and complex dust geometry.

In the top panel of Fig. \ref{fig:irx}, we present the IRX--$\buv$ relation for our simulated sample, color-coded by redshift (points; $\fdust=0.4$). $\FIR$ is the total dust flux integrated over 8--1000\,$\mu$m and $\FUV$ is the neutral flux density, $\lambda F_{\lambda}$, at {1500\,\AA}. $\buv$ is measured using the monochromatic flux at two wavelengths {1500\,\AA} and {2300\,\AA}. We only show one sightline for each galaxy snapshot, but highlight all five lines of sight for galaxies z5m12b and z9m12a at $z=5$ and 9, respectively (example images shown in Fig. \ref{fig:image}), using blue and orange circles to illustrate the variation from different viewing angles. We compare our results with the observational dataset compiled in \citet{Casey:2018a}, which consists of the ASPECS-Pilot sample from \citet{Aravena:2016} and $z\sim5.5$ sample from \citet{Capak:2015} with updated measurements by Barisic et~al. and Casey et~al. We also show the empirical IRX--$\buv$ relation developed from local starburst galaxies in \citet[][dashed]{Meurer:1999} and the aperture-corrected relation in \citet[][dotted]{Takeuchi:2012}. The solid lines show the IRX--$\buv$ relation derived from simple dust screen model applying a SMC-like attenuation law \citep[e.g.][]{Pettini:1998}.

The simulated galaxies form a tight IRX--$\buv$ relation. There is a small redshift evolution with galaxies moving toward bluer $\buv$ at fixed IRX with increasing redshift, simply because of younger stellar populations at higher redshifts \citep[see also e.g.][]{Grasha:2013}. Line-of-sight variations for the same galaxy follow the sample IRX--$\buv$ relation. Our simulated sample broadly lies within the scatter of recent measurements and agrees well with the SMC IRX--$\buv$ relation derived from a simple dust screen picture. Surprisingly, although our simulations show patchy, complex dust distribution in these galaxies and the radiative transfer calculations reveal non-trivial effects of dust scattering in the rest-frame UV (see Fig. \ref{fig:image}), we suggest that they do not drive significant scatter in the IRX--$\buv$ relation.

In the bottom panel of Fig. \ref{fig:irx}, we compare the IRX--$\buv$ relation for different dust-to-metal ratios (color points). Note that this must be understood as varying the normalization of the extinction curve. We only show a subsample of our galaxy snapshots (at integer redshifts) along the same line of sight as in the top panel. We find that the IRX--$\buv$ relation does not change with $\fdust$. We also examine the situation where the UV and IR fluxes are measured using a smaller aperture ($\Rmax/3$ instead of $\Rvir$; red squares) and find it has no significant effect on the IRX--$\buv$ relation. Finally, as a proof of concept, we did an experiment where we only consider dust extinction along the line of sight using the total extinction opacity in Fig. \ref{fig:opacity} but ignore dust scattering (grey squares). This effectively changes the attenuation law. The sample is more consistent with the Calzetti-like IRX--$\buv$ relation by coincidence, as the extinction opacity is shallower than the absorption opacity in the UV (see Fig. \ref{fig:image}). Our results suggest that the IRX--$\buv$ relation is mainly determined by the shape of the extinction curve and independent of its normalization, at least in the mass range we probe in our simulations. If the large scatter in the IRX--$\buv$ relation reported in some $z\geq5$ galaxy sample is real, it is more likely to be caused by variations in the effective attenuation law, rather than by smaller dust-to-metal ratios in high-redshift galaxies.

\subsection{Bolometric IR luminosity}
\label{sec:bolIR}
In this section, we present the bolometric luminosity of dust emission ($\LIR$, integrated over 8--1000\,$\mu$m) for our simulated sample and its dependence on various galaxy properties. We include all the light within $\Rvir$ to calculate $\LIR$, which is not a bad treatment as instruments probing these wavelengths usually have large beam sizes. These results are not only useful for understanding the physics of dust obscuration and emission in high-redshift galaxies, but also important for empirically modeling the abundances of dusty star forming galaxies at these redshifts.

The left panel of Fig. \ref{fig:lirluv} shows the relationship between dust bolometric luminosity in the IR, $\LIR$, and galaxy {\em intrinsic} UV luminosity (prior to dust attenuation), $\LUVi\equiv\lambda L_{\lambda}$ at 1500\,{\AA} (including all the light within $\Rvir$). Each point represents one galaxy snapshot, color-coded by redshift. There is a correlation between $\LIR$ and $\LUVi$ that can be well described by a broken power-law function
\be
\label{eqn:lirluv}
 \LIR = \frac{L_{\rm IR}^{\ast}}{\left(\frac{\LUVi}{L_{\rm UV,\,intr}^{\ast}}\right)^{\gamma_1} + \left(\frac{\LUVi}{L_{\rm UV,\,intr}^{\ast}}\right)^{\gamma_2}} ,
\ee
where $(\gamma_1,\gamma_2,L_{\rm UV,\,intr}^{\ast},L_{\rm IR}^{\ast}) = (-1.23, -1.86, 10^{9.81}, 10^{9.26})$ for $\fdust=0.4$ (the red dashed line in the middle). This suggests that dust attenuation and emission become weaker (from $\sim\LUVi^{1.23}$ to $\sim\LUVi^{1.86}$) below $\LUVi\sim10^{10}\,\Lsun$ (i.e. $\sim-19$\,mag). We find that $\LIR$ is roughly comparable to $\sim60$--70\% of the bolometric luminosity at the most luminous end of our sample (note $\LUVi=0.68\,\Lbol$). This is consistent with, but slightly lower than those (60--90\%) found in \cite{Cen:2014}.

We also show how the bolometric IR luminosity changes with respect to the normalization of the extinction curve (represented by varying $\fdust$). The dash-dotted and dotted lines in the left panel of Fig. \ref{fig:lirluv} show the broken power-law fits for $\fdust=0.2$ and 0.8, respectively. The best-fit parameters $(\gamma_1,\gamma_2,L_{\rm UV,\,int}^{\ast},L_{\rm IR}^{\ast})$ are 
\begin{eqnarray*}
& & (-1.28, -1.86, 10^{10.00}, 10^{9.32}) ~ \text{for} ~ \fdust=0.2 ~ \text{and} \\ 
& & (-1.20, -1.85,~ 10^{9.56}, 10^{9.08}) ~ \text{for} ~ \fdust=0.8.
\end{eqnarray*}
At the bright end, the bolometric dust luminosity does not change with $\fdust$ by more than 0.1\,dex, because most of the obscured sightlines are optically thick. At the faint end where the galaxies are optically thin to dust attenuation, $\LIR$ is proportional to $\fdust$. 

Galaxies at $z=5$--12 lie on the same $\LIR$--$\LUVi$ relationship. In the right panels of Fig. \ref{fig:lirluv}, we explore why there is no redshift dependence on the $\LIR$--$\LUVi$ relation and what drives its scatter. To this end, we take all simulated galaxies in a narrow range of UV luminosity (within 0.1\,dex from $\LUVi=10^{11}\,\Lsun$, as labeled by the grey rectangular in the left panel of Fig. \ref{fig:lirluv}) and search for secondary dependence of $\LIR$ on total dust mass ($\Mdust$, top left), average dust surface density ($\langle\Sigma\rangle_{\rm dust}\equiv\Mdust/(\Rmax/3)^2$, top right), and average dust density ($\langle\rho\rangle_{\rm dust}\equiv\Mdust/(\Rmax/3)^3$, bottom left).\footnote{Here we adopt $\Rmax/3$ as a characteristic size of dust distribution mainly for illustrative purposes. Other size measures, such as half-mass/half-light radius, are statistically scaled with $\Rmax$ up to a constant factor.} Interestingly, all three quantities are redshift-dependent as expected: at fixed $\LUVi$, galaxies at higher redshifts contain less dust mass\footnote{As shown in Fig. \ref{fig:smhm}, $\LUVi$ increases with redshift at fixed halo mass and stellar mass from $z=5$--12 \citep[see also][]{Ma:2018a}. Therefore, at fixed $\LUVi$, galaxies at higher redshifts are less massive and thus less dust rich.} but show higher average dust column density (equivalent to optical depth) and density. However, {\em none} of these quantities correlate with $\LIR$. This suggests that at a given $\LUVi$, the dust luminosity is primarily determined by the covering fraction of optically-thick sightlines, regardless of total dust mass and  dust density in the system.

The bottom right panel of Fig. \ref{fig:lirluv} shows that the scatter of the $\LIR$--$\LUVi$ relation is driven by the SFR averaged over the past 10\,Myr, in other words, the amount of stars younger than 10\,Myr in a galaxy\footnote{Note that there is a correlation between $\SFR_{\rm 10\,Myr}$ and $\LUV$, but they are not fully degenerate: stars older than 10\,Myr still provide a non-negligible fraction of the UV light, depending on the recent star formation history.}. Note that if the SFR is measured over longer time-scale (e.g. 100\,Myr), the secondary dependence of $\LIR$ on SFR at fixed $\LUV$ becomes weaker. The physical picture behind this result is that stars younger than 10\,Myr are {\em more heavily obscured} by their birth cloud than relatively older stars (e.g. 10--100\,Myr, which still contribute a significant fraction of the UV light). This is consistent with models where differential obscuration between young stars and older stars is applied by hand \citep[e.g.][]{Charlot:2000,Jonsson:2010,Katz:2018}, albeit our simulations explicitly resolve this with our star formation and feedback models.

\begin{figure}
\centering
\includegraphics[width=\linewidth]{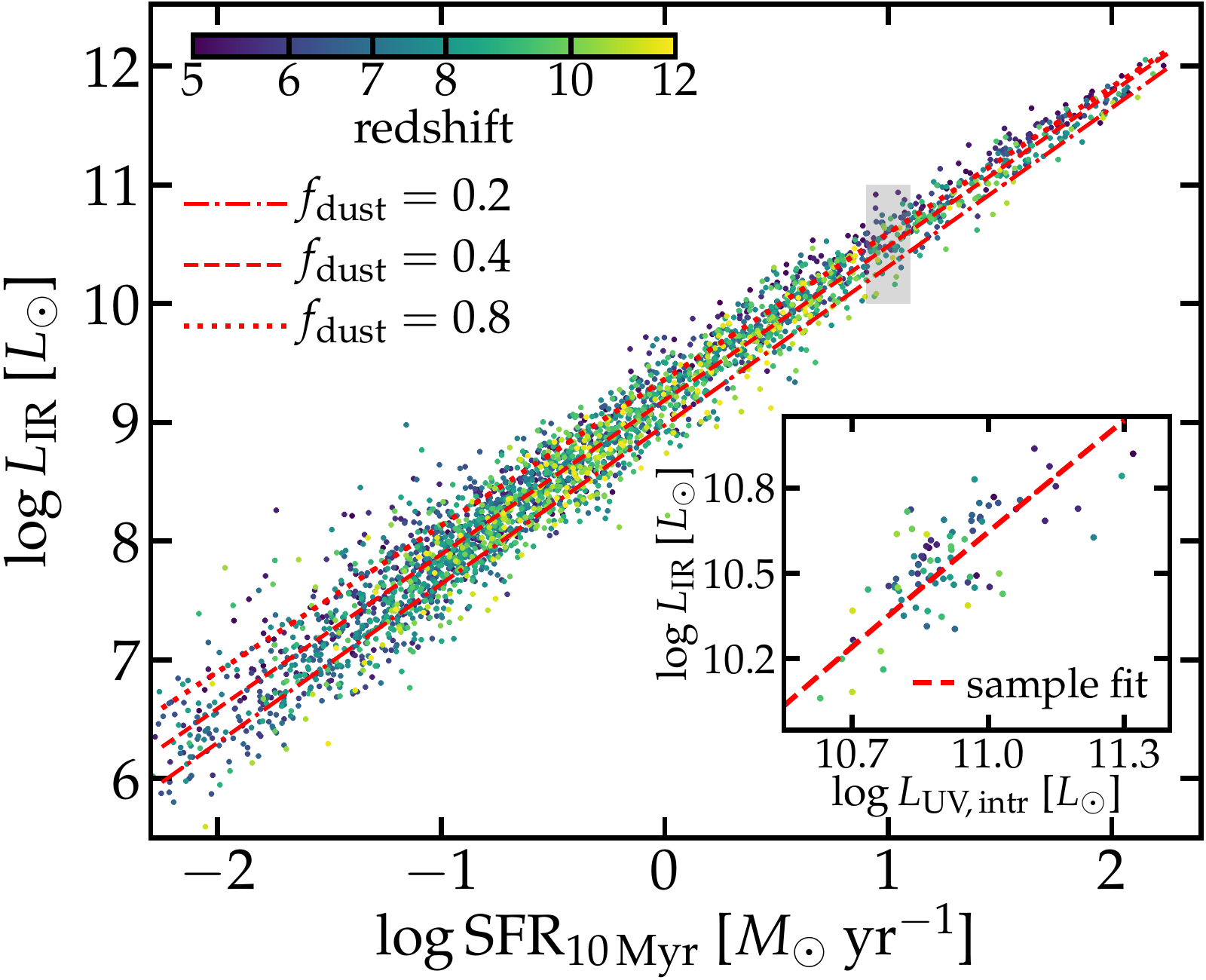}
\caption{The $\LIR$--$\SFR_{\rm 10\,Myr}$ relation. Each point represents a galaxy snapshot in our sample, color-coded by redshift (using $\fdust=0.4$). This relation is best described by a power-law function (Equation \ref{eqn:lirsfr}, red dashed line). The dependence on $\fdust$ is shown by the three red lines. Using galaxies at fixed $\SFR_{\rm 10\,Myr}$ (0.1\,dex from $10\,\Msun\,\yr^{-1}$, the grey shaded region), the smaller panel shows that the scatter in the $\LIR$--$\SFR_{\rm 10\,Myr}$ relation is driven by $\LUVi$. This means that differential obscuration for young and relatively older stars is important for understanding dust attenuation and emission.}
\label{fig:lirsfr}
\end{figure}

Fig. \ref{fig:lirsfr} shows the $\LIR$--$\SFR_{\rm 10\,Myr}$ relation for the entire simulated sample. Each point represents one galaxy snapshot, color-coded by redshift. Again, the $\LIR$--$\SFR_{\rm 10\,Myr}$ relationship does not depend on redshift (as well as dust mass and density as we explicitly checked). This relation is best described by a single power-law function
\be
\label{eqn:lirsfr}
\log\LIR = \gamma \, \log \left( \frac{\SFR_{\rm 10\,Myr}}{1\,\Msun\,\yr^{-1}} \right) + \delta,
\ee
where $(\gamma,\delta)=(1.30,9.19)$ for $\fdust=0.4$ as shown by the red line in the main panel of Fig. \ref{fig:lirsfr} (as well as in the bottom right panel of Fig. \ref{fig:lirluv}). The $\LIR$--$\SFR_{\rm 10\,Myr}$ relation also changes with $\fdust$ as $(\gamma,\delta)$ $=(1.34,8.98)$ for $\fdust=0.2$ and (1.23,9.36) for $\fdust=0.8$.

The smaller panel at the bottom right corner of Fig. \ref{fig:lirsfr} shows the secondary dependence of $\LIR$ on $\LUVi$ at fixed $\SFR_{\rm 10\,Myr}$ (within 0.1\,dex from $10\,\Msun\,\yr^{-1}$, as labeled by the grey rectangular in the main panel). The red dashed line shows the double power-law fit in the left panel of Fig. \ref{fig:lirluv}. This is because stars older than 10\,Myr provide an extra source for dust emission. Combining the results in Figs. \ref{fig:lirluv} and \ref{fig:lirsfr} further confirms that differential obscuration between young and relatively old stars is important in understanding dust attenuation and emission. Finally, we inspect the galaxies at both fixed $\LUV$ and $\SFR_{\rm 10\,Myr}$ and find no further dependence of $\LIR$ on other dust properties: this is the intrinsic scatter purely due to variations of dust geometry in these galaxies.

\subsection{Dust SEDs and dust temperature}
\label{sec:sed}
In this section, we study the dust SEDs and dust temperatures for our simulated sample. Again, we include all the light in $\Rvir$. As we show in Fig. \ref{fig:image}, there is a broad distribution of dust temperatures in a single galaxy, with dust close to the young stars being heated up to 100\,K and diffuse dust at large radii at much lower temperatures. It is thus non-trivial to parametrize dust SEDs and even define one dust `temperature'. One of the most commonly adopted forms for modeling dust SEDs is the MBB function for single-temperature dust \citep[e.g.][]{da-Cunha:2013,Bouwens:2016}
\be
\label{eqn:mbb}
    L_{\nu} \propto (1-e^{-\tau_{\nu}}) \, B_{\nu} (T) \sim \nu^{\beta} B_{\nu}(T) = \frac{\nu^{3+\beta}}{e^{h\nu/kT}-1},
\ee
where the second expression is valid in the optically thin limit and a power-law opacity $\kappa_{\nu}\propto\nu^{\beta}$ is applied. In this situation, the peak wavelength of $L_{\nu}$\footnote{Note that the peak wavelengths of $L_{\nu}$, $L_{\lambda}$, and $\nu L_{\nu}$ are different.} is $\lp=96.64\,\mu{\rm m}\,(30\,{\rm K}/T)$ and the total dust luminosity is $\propto T^{4+\beta}$. A more realistic form is the two-component dust SED model, consisting of a MBB function for old dust and a power-law component for warmer dust \citep{Casey:2012}. Nevertheless, an optically-thin MBB function at local equilibrium temperature is still a good approximation for the local dust emissivity at rest-frame $\lambda>30\,\mu$m where NLTE effects are negligible (see Section \ref{sec:rt}).

We adopt three definitions of dust temperature that we will refer to in the discussion below. First, we define the peak temperature $\Tp = 30\,{\rm K} \, ( 96.64\,\mu{\rm m}/\lp )$. Note that $\Tp$ is only a proxy for $\lp$, so the normalization here is just a choice of ours, which is adopted from the peak wavelength $\lp$ of $L_{\nu}$ for an optically-thin MBB function. Next, we introduce the mass-weighted dust temperature 
\be
    T_{\rm mw} = \int T_{\rm eq} \, \rho_{\rm dust} \, \dd V \bigg/  \int \rho_{\rm dust} \, \dd V,
\ee
where $T_{\rm eq}$ is the equilibrium dust temperature given by {\sc skirt} under the LTE assumption (see Equation \ref{eqn:lte}). This is the most straightforward one to calculate from dust radiative transfer calculations for simulated galaxies and adopted by various authors in the literature (e.g. \citealt{Behrens:2018}; Liang et al. 2019). It is worth noting that the mass-weighted temperature directly relates to the dust SED at the R--J tail (e.g. \citealt{Scoville:2016}; Liang et al. 2019). Finally, we define the effective dust temperature 
\be
    \Teff = \left(  \int T_{\rm eq}^{4+\beta} \rho_{\rm dust} \, \dd V \bigg/ \int \rho_{\rm dust} \, \dd V \right)^{\frac{1}{4+\beta}} .
\ee
Note that the frequency-integrated dust emissivity (power per unit volume) is $\int j_{\nu}\,\dd\nu=\int \alpha_{\nu}\,B_{\nu}(T)\,\dd\nu=\int \kappa_{\nu,\,\rm dust}\,\rho_{\rm dust}\,B_{\nu}(T)\,\dd\nu \sim \int \nu^{\beta} \,\rho_{\rm dust}\,B_{\nu}(T)\,\dd\nu \propto T^{4+\beta} \rho_{\rm dust}$ ($j_{\nu}$ is emissivity and should not be confused with the radiation intensity $J_{\nu}$ in Equation \ref{eqn:lte}), so the effective dust temperature is defined such that in the optically thin limit, the bolometric dust luminosity is $\LIR\propto \Mdust \Teff^{4+\beta}$.\footnote{Again, we remind that dust opacity is degenerate with dust-to-metal ratio, as it is always $\kdust\fdust$ that appears in the absorption coefficient. Therefore, if we fix $\fdust$ at 0.4 but boost $\kdust$ by a factor of 2, the radiation field and dust temperature will remain identical to our $\fdust=0.8$ calculation at fixed $\kdust$. For simplicity, we use $\Mdust$ to interpret the results for different $\fdust$ below, but one should note that in also includes the uncertainty of dust opacity.} All three temperatures correlate with each other with large scatter depending on the exact dust temperature distribution in each galaxy.

\begin{figure}
\centering
\includegraphics[width=\linewidth]{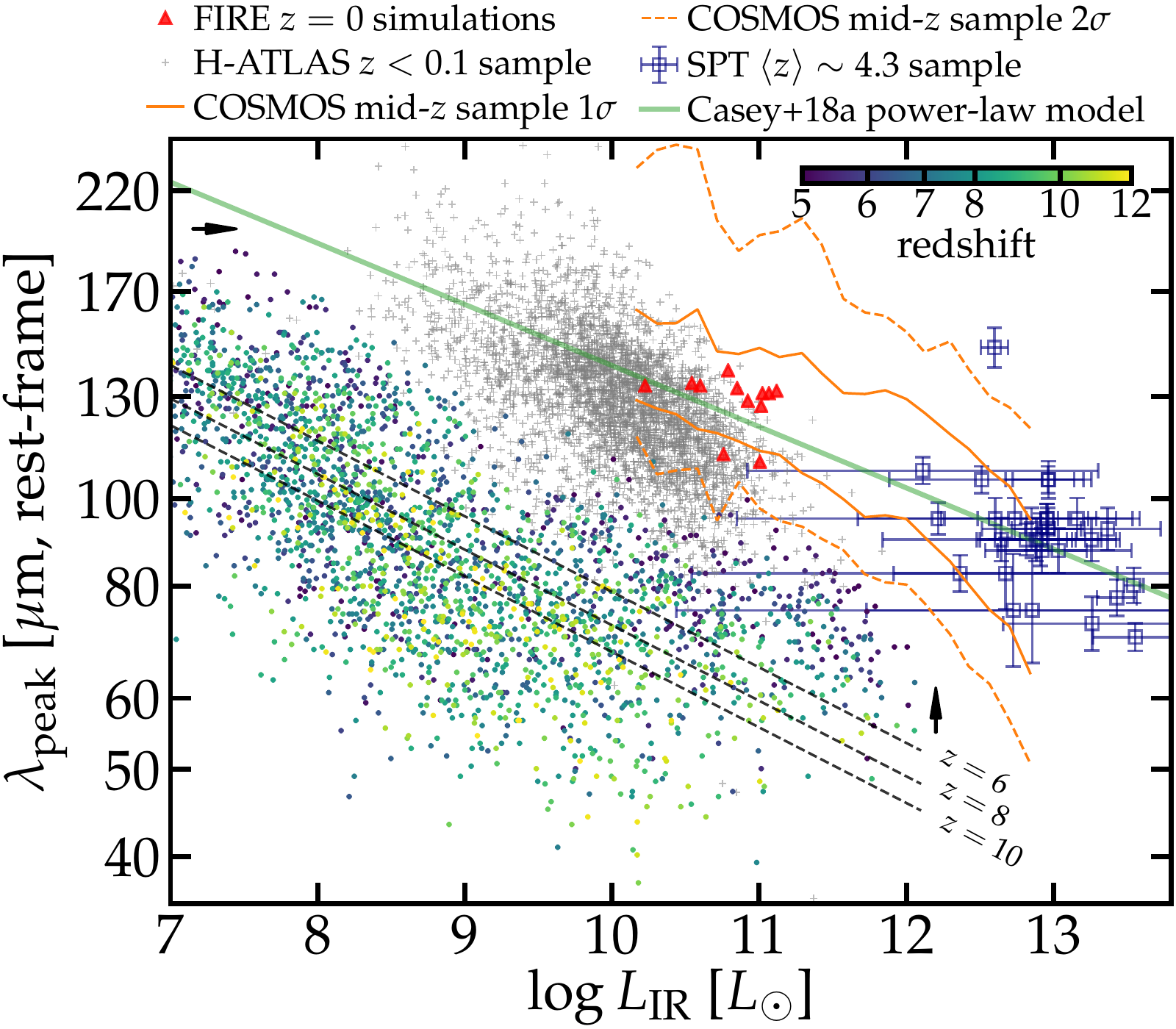}
\caption{The relation between peak wavelength $\lp$ of $L_{\nu}$ and bolometric IR luminosity $\LIR$. Each color point represents one galaxy snapshot in our $z\geq5$ simulation sample, color-coded by redshift (using $\fdust=0.4$). The red triangles show the 12 MW-mass galaxy simulations at $z=0$ from the FIRE suite run with the same code and comparable resolution. We compare with the observational data compiled in \citet{Casey:2018}, including the low-redshift H-ATLAS sample (\citealt{Valiante:2016}; all $z<0.1$ galaxies shown by grey points), the intermediate-redshift COSMOS sample (\citealt{Lee:2013}; $0.5<z<2$, $1\sigma$ and $2\sigma$ ranges shown by orange lines), and the SPT-detected $\langle z\rangle\sim4.3$ DSFG sample (\citealt{Strandet:2016}; blue squares). The green line shows the best-fit power-law model in \citet{Casey:2018}. With our $z=0$ simulations in good agreement with observation, we predict that the $z\geq5$ sample peaks at a factor of 2 shorter wavelengths (indicating higher dust temperatures) than low-redshift galaxies at the same $\LIR$. The black dashed lines show the best-fit power-law function $\lp=78.78\,\mu{\rm m} \, [(1+z)/7]^{-0.34}\,(\LIR/10^{10}\,\Lsun)^{-0.084}$ (cf. Equation \ref{eqn:tlir}). The arrows show how the faintest and brightest $z\geq5$ galaxies move on the $\lp$--$\LIR$ plane, respectively, if $\fdust$ increases by a factor of 2.}
\label{fig:lirlpeak}
\end{figure}

\cite{Casey:2018} suggested a redshift-independent, empirical relation between observed rest-frame peak wavelength $\lp$ of $L_{\nu}$ and bolometric IR luminosity $\LIR$ derived from several observed samples from $z=0$--6. These include data from the H-ATLAS survey mostly covering $0<z<0.5$ \citep{Valiante:2016}, the sample in the COSMOS field at $0.3<z<2$ with {\it Herschel} detection \citep{Lee:2013}, and the South Pole Telescope (SPT)-detected DSFGs sample with average redshift $\langle z\rangle\sim4.3$ from \cite{Strandet:2016} and \citet{Spilker:2016}. Both $\lp$ and $\LIR$ were re-measured by fitting the two-component dust SED model to the original data. In Fig. \ref{fig:lirlpeak}, we compile the individual galaxies in the H-ATLAS $z<0.1$ sample (grey points), $1\sigma$ and $2\sigma$ ranges for the COSMOS sample (orange solid and dashed lines), the SPT $\langle z\rangle\sim4.3$ DSFG sample (blue squares), and the best-fit power-law model 
\be
\label{eqn:lirlpeak}
    \lp = 102.8\,\mu{\rm m}\,\left(\frac{\LIR}{10^{12}\,\Lsun}\right)^{-0.068} 
\ee
(the green line), all taken from \cite{Casey:2018}, for comparing with our simulations.

In Fig. \ref{fig:lirlpeak}, we also present the $\lp$--$\LIR$ relation for all galaxy snapshots at $z>5$ from our simulated sample (color points; using $\fdust=0.4$). For a sanity check, we conduct dust radiative transfer calculations using identical methods on a sample of 12 Milky Way (MW)-mass galaxies from the FIRE simulations at $z=0$, including 8 isolated halos and 2 Local Group (LG)-like galaxy pairs at mass resolution $\mb=3500$--$7000\,\Msun$\footnote{Among these simulations, 6 isolated halos and 2 LG-like pairs (10 galaxies) have been presented in \citet{Garrison-Kimmel:2018}. The other two isolated halos will be presented in Garrison-Kimmel et al. (in preparation).} (comparable or better than those studied in this paper), run with the identical version of {\sc gizmo}. The $\lp$--$\LIR$ relation for the $z=0$ FIRE sample consisting of 12 MW-mass galaxies is shown by the red triangles in Fig. \ref{fig:lirlpeak}.

The FIRE simulations at $z=0$ agree well with the observed $\lp$--$\LIR$ relation for the H-ATLAS $z<0.1$ sample and lies along the empirical power-law relation from \citet[][Equation \ref{eqn:lirlpeak}]{Casey:2018}. However, although the $z\geq5$ sample also shows an anti-correlation between $\lp$ and $\LIR$, it is offset from the observational data and the $z=0$ simulations, with $\lp$ moving toward shorter wavelengths by a factor of 2 at a given $\LIR$. At the most luminous end in our sample, we find $\lp\sim60$--80\,$\mu$m, in good agreement with previous simulations at similar redshifts post-processed with dust radiative transfer calculations \citep[e.g.][]{Cen:2014}. This suggests that dust is much warmer in $z>5$ galaxies than in low-redshift galaxies. In fact, the effective dust temperature in our $z=0$ MW-mass galaxy simulations is $\sim18$\,K, whereas it is typically over 35\,K in the $z>5$ galaxies at similar IR luminosities ($\LIR=10^{10}$--$10^{11}\,\Lsun$). Given that our $z=0$ simulations are in good agreement with observations, we argue that this prediction is a physical effect, as the $z=0$ and $z>5$ simulation samples are run with the same code and comparable resolution.\footnote{We have also checked the progenitors of the 12 MW-mass galaxies at $z>0$ and found that they lie between the $z=0$ and the $z\geq5$ samples on the $\lp$--$\LIR$ relation, with $\lp$ decreasing with redshift at fixed $\LIR$.} The black dashed lines in Fig. \ref{fig:lirlpeak} show the best-fit power-law function of the $\lp$--$\LIR$ relation for our sample, $\lp=78.78\,\mu{\rm m} ~ [(1+z)/7]^{-0.34} ~ (\LIR/10^{10}\,\Lsun)^{-0.084}$, at $z=6$, 8, and 10 (see Equation \ref{eqn:tlir} and Table \ref{tbl:tlir} for details).

The SPT-detected DSFG sample at $\langle z\rangle\sim4.3$ seems to lie on the same $\lp$--$\LIR$ relation as low-redshift galaxies. This does not necessarily mean that the $\lp$--$\LIR$ relation is redshift independent out to $z\geq5$. First, the SPT-detected galaxies are much more luminous than our simulated galaxies. More important, galaxies of similar $\LIR$ but shorter $\lp$ have weaker flux densities at long wavelengths where the observations are conducted, so they tend to be excluded in a flux-limited sample. Therefore, the SPT sample cannot falsify our prediction that most $z\geq5$ galaxies have a factor of 2 shorter $\lp$ than lower-redshift galaxies.

\begin{figure}
\centering
\includegraphics[width=\linewidth]{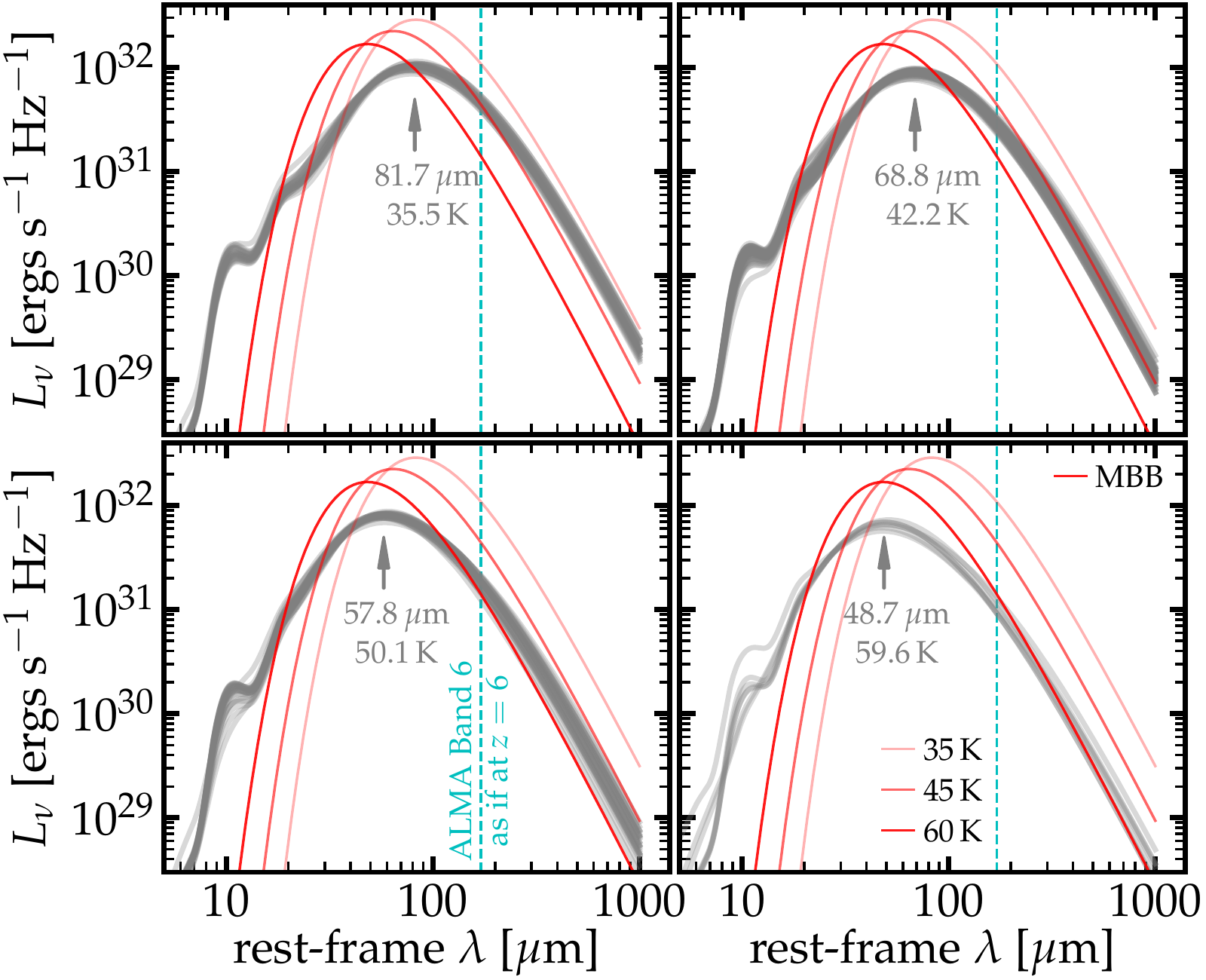}
\caption{Rest-frame dust SEDs for our $z>5$ simulations (using $\fdust=0.4$). In each panel, we show all galaxies brighter than $\LIR=10^{10}\,\Lsun$ with $\lp$ falling in a 0.05\,dex bin centered on the wavelength marked by the grey arrow. All SEDs are renormalized to $\LIR=10^{11.5}\,\Lsun$. The red lines in each panel show the optically-thin MBB function at $T=35$, 45, and 60\,K, also normalized to $10^{11.5}\,\Lsun$. The vertical cyan dashed lines illustrate the observed-frame 1.2\,mm (ALMA Band 6) for $z=6$ (rest-frame 171\,$\mu$m). A 35\,K MBB function overestimates the flux density at this wavelength by a factor of 3--10. To convert between $\LIR$ and observed-frame 1.2\,mm flux density for $z\sim6$ galaxies using an optically MBB function, one must adopt a high dust temperature of 45--60\,K.}
\label{fig:sed}
\end{figure}

The black arrows in Fig. \ref{fig:lirlpeak} indicate the amount and direction that faintest and brightest galaxies in our sample move along on the $\lp$--$\LIR$ plane, respectively, if $\fdust$ increases by a factor of 2 (i.e. $\fdust=0.8$). At the faint end, the total dust luminosity increases by a factor of 2 (see Fig. \ref{fig:lirluv}) while the dust mass also doubles. We thus expect the dust effective temperature (so does the peak temperature or $\lp$) remains unchanged, so faint galaxies move horizontally to higher $\LIR$ by approximate 0.3\,dex. At the bright end, the total dust luminosity changes very little, so the effective temperature should decrease by a factor of $2^{1/(4+\beta)}=1.12$. Therefore, bright galaxies move vertically toward longer $\lp$ by nearly 0.05\,dex. This brings our $z\geq5$ sample closer to the observed $\lp$--$\LIR$ relation, although a large offset still remains. Following a similar argument, galaxies will move along the opposite direction by the same distance shown by the arrows if $\fdust$ decreases by a factor of 2. This is confirmed by our radiative transfer calculations.

\begin{figure*}
\centering
\includegraphics[width=\linewidth]{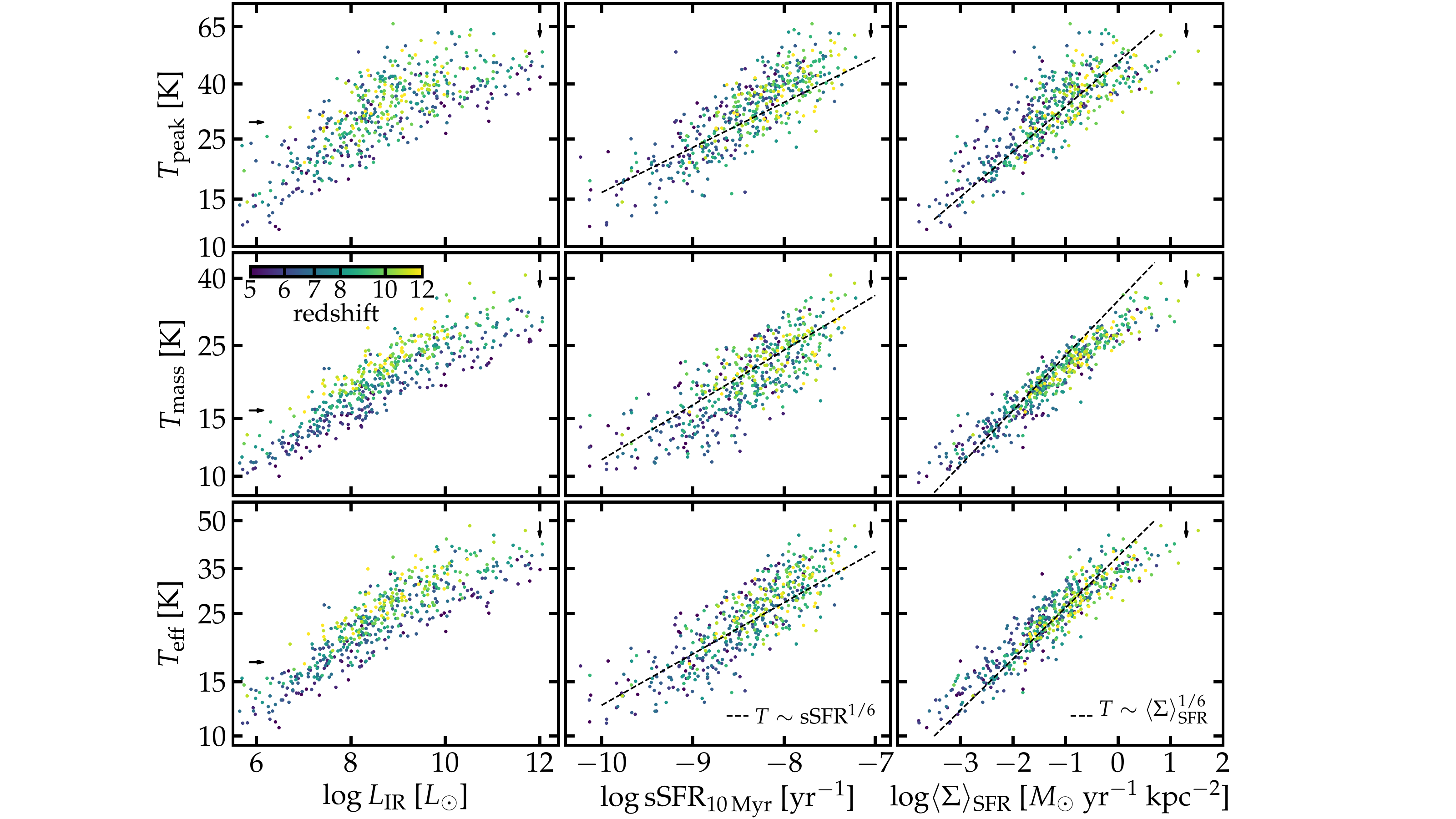}
\caption{The correlation between dust temperature and bolometric IR luminosity (left), specific star formation rate (middle, averaged over the past 10\,Myr), and SFR surface density ($\langle\Sigma\rangle_{\rm SFR}\equiv{\rm SFR_{10\,Myr}}/(\Rmax/3)^2$, right). Each row represents one definition of dust temperature, including peak temperature (top), mass-weighted temperature (middle), and effective temperature (bottom). We only show snapshots at integer redshifts, color-coded by redshift ($\fdust=0.4$). The dust temperature--sSFR correlation can be understood as $T\sim(\LIR/\Mdust)^{1/6}\sim(\SFR/\Ms)^{1/6}\equiv{\rm sSFR}^{1/6}$, given that $\LIR\sim\SFR^{1.3}$ and $\Mdust\sim\Ms$. At fixed $\LIR$, dust temperature increases with redshift, as galaxies tend to have higher sSFR at higher redshifts. $\langle\Sigma\rangle_{\rm SFR}$ reflects the intensity of interstellar radiation on dust grains ($\sim L/R^2$), which sets the dust temperature by $T\sim\langle\Sigma\rangle_{\rm SFR}^{1/6}$. The black dashed lines illustrate the $1/6$-power scaling relations as argued above, which are in broad agreement with the simulations. The black arrows show how galaxies at the faint/bright end move if $\fdust$ increases by a factor of 2. The best-fit redshift-dependent $T$--$\LIR$ relation (Equation \ref{eqn:tlir}) for $\fdust=0.2$, 0.4, and 0.8 and for all dust temperature definitions are given in Table \ref{tbl:tlir}.}
\label{fig:Tcomp}
\end{figure*}

In Fig. \ref{fig:sed}, we present the rest-frame dust SEDs for our $z\geq5$ simulations (using $\fdust=0.4$). In each panel, we collect all galaxies brighter than $\LIR=10^{10}\,\Lsun$ with peak wavelength falling in a 0.05\,dex bin around the wavelength marked by the grey arrow (both $\lp$ and $\Tp$ are labeled in each panel). The median $\lp$ differs by 0.075\,dex between adjacent panels. The shape of dust SED in mid- and far-IR does not strongly depend on $\LIR$ and redshift for galaxies in such narrow bins of $\lp$, so we rescale all galaxies to $\LIR=10^{11.5}\,\Lsun$. In each panel, we also show optically-thin MBB functions (Equation \ref{eqn:mbb}) at $T=35$, 45, and 60\,K (all normalized to $10^{11.5}\,\Lsun$; red lines) for reference. The vertical cyan dashed lines in Fig. \ref{fig:sed} label the observed-frame 1.2\,mm (ALMA Band 6) at $z=6$ (rest-frame 171\,$\mu$m). For our $z\geq5$ galaxies, their flux densities at this wavelength are comparable to $T\sim45$--60\,K MBB emission at the same $\LIR$. For $\fdust=0.8$, the peak wavelengths increases and characteristic temperatures decrease by a factor of $2^{1/6}=1.12$. The mid-IR SED (around $\lp$) is shaped by warm dust, which is not accounted for by the single-temperature MBB function. Note that at rest-frame $\lambda=6$--25\,$\mu$m, the SED is usually dominated by PAH line emission \citep[e.g.][]{Baes:2011}, which is not included in the \cite{Weingartner:2001} SMC dust model. Our results at these wavelengths should be used with caution in this regard.

\cite{Bouwens:2016} find that the ALMA Band 6 (observed-frame 1.2\,mm continuum) deep survey in the HUDF detects much fewer $\sim L^{\ast}$ galaxies at $z>4$ than what inferred from galaxy rest-frame UV slopes, even assuming the shallower SMC IRX--$\buv$ relation, {\em unless} dust temperature increases to 44--50\,K (as oppose to 35\,K) at $z>4$ (such that their far-IR fluxes are too weak to detect). They first derive the total IR luminosities of UV-selected galaxies from their UV fluxes and $\buv$ and then convert $\LIR$ to observed-frame 1.2\,mm fluxes assuming the dust SEDs follow optically-thin MBB functions of assumed temperatures. Our simulated galaxies follow the SMC IRX--$\buv$ relation, but a 35\,K MBB function overestimates the flux density at ALMA Band 6 wavelength by a factor of 3--10. In other words, to convert between $\LIR$ and observed-frame 1.2\,mm flux for $z\sim6$ galaxies using an optically-thin MBB function, one must assume a dust temperature of 45--60\,K. These results support the hypothesis that the low detection rate of high-redshift galaxies in mm surveys is caused by galaxies falling below the detection limit because of their high dust temperatures \citep[see also][]{Faisst:2017}. The existence of an IRX--$\buv$ relation close to the local or the SMC relation in $z>5$ galaxies cannot be ruled out. 

\begin{table}
\caption{Best-fit parameters for the redshift-dependent $T$--$\LIR$ relation (see Equation \ref{eqn:tlir}) for $\fdust=0.2$, 0.4, and 0.8, in ($T_0$,\,$a_1$,\,$a_2$).} 
\begin{tabular}{cccc}
\hline
$\fdust$ & $T_{\rm mw}$ & $T_{\rm eff}$ & $T_{\rm peak}$ \\
\hline
0.2 & (24.0,\,0.40,\,0.078) & (30.6,\,0.37,\,0.087) & (40.7,\,0.34,\,0.089) \\
0.4 & (22.5,\,0.41,\,0.076) & (28.2,\,0.37,\,0.084) & (36.8,\,0.34,\,0.084) \\
0.8 & (20.9,\,0.41,\,0.073) & (26.0,\,0.37,\,0.081) & (33.3,\,0.33,\,0.079) \\
\hline
\end{tabular}
\label{tbl:tlir}
\end{table}

In Fig. \ref{fig:Tcomp}, we present the correlation between dust temperature and bolometric IR luminosity $\LIR$ (left), specific star formation rate (over the past 10\,Myr; middle), and average SFR surface density ($\langle\Sigma\rangle_{\rm SFR}\equiv{\rm SFR_{10\,Myr}}/(\Rmax/3)^2$, right)\footnote{Again, $\Rmax/3$ is adopted here as a characteristic scale. Other size measures are expected to scale up to a constant factor in a statistical sense.}. Each row shows one definition of dust temperature, with $\Tp$ in the top, $T_{\rm mw}$ in the middle, and $\Teff$ in the bottom. We only show a subsample of snapshots, color-coded by redshift. All the three temperatures correlate with $\LIR$, consistent with the negative $\lp$--$\LIR$ correlation shown in Fig. \ref{fig:lirlpeak}. At the same $\LIR$, dust temperature increases with redshift. We fit the $T$--$\LIR$ relation for our simulated sample by the redshift-dependent power-law function
\be
\label{eqn:tlir}
T = T_0  \left( \frac{1+z}{7} \right)^{a_1} \left( \frac{\LIR}{10^{10}\,\Lsun} \right)^{a_2}.
\ee
We list the best-fit parameters for $T_{\rm mw}$, $T_{\rm eff}$, and $T_{\rm peak}$ and for $\fdust=0.2$, 0.4, and 0.8 in Table \ref{tbl:tlir}. Note that these fitting functions should only apply to star-forming galaxies below $\LIR\sim10^{12}\,\Lsun$ at $z=5$--12. Dust temperatures in $z\sim2$--4 DSFGs are studied in more detail in Liang et al. (2019) using a separate suite of FIRE simulations.

In contrast, neither the $T$--sSFR nor the $T$--$\langle\Sigma\rangle_{\rm SFR}$ relation depends on redshift. Here we want to provide simple, qualitative understanding on these correlations first, so we do not distinguish the three dust temperatures defined above for simplicity, although they are conceptually and physically different (see e.g. Liang et al. 2019 for more details). The $T$--sSFR relation can be understood, given $\LIR\propto\SFR^{1.3}$ (Fig. \ref{fig:lirsfr}) and $\Mdust\propto\Ms$ (Fig. \ref{fig:smhm}), as \citep[see also][]{Magnelli:2014,Safarzadeh:2016}
\be
    \Teff \propto \left( \frac{\LIR}{\Mdust} \right)^{1/6} \propto {\rm sSFR}^{1/6} \cdot \SFR^{0.05} ~ ({\rm for}~\beta=2),
\ee
where the second term is subdominant. As we have shown in \cite{Ma:2018a}, SFR increases with redshift at fixed stellar mass from $z=5$--12 (see also Fig. \ref{fig:smhm}). The redshift-dependence of the $T$--$\LIR$ relation can thus be attributed to the increasing sSFR with redshift \citep[i.e. luminosity per unit dust mass; see also][]{Imara:2018}. The $T$--$\langle\Sigma\rangle_{\rm SFR}$ relation is probably more physically expected. Given that SFR is proportional to the total luminosity from stellar sources and $\Rmax/3$ is a characteristic scale of dust distribution, $\langle\Sigma\rangle_{\rm SFR}$ reflects the intensity of the interstellar radiation field on dust grains (i.e. $\langle\Sigma\rangle_{\rm SFR}\sim L/R^2\sim J$), which sets the dust temperature via energy balance. Following Equation \ref{eqn:lte}, $T$ should also scale to the $1/6$ power of $\langle\Sigma\rangle_{\rm SFR}$. The black dashed lines in Fig. \ref{fig:Tcomp} show the power-law scaling relations derived from the simple arguments above, in broad agreement with our simulated sample. Note that at a given luminosity, galaxies tend to be more compact at higher redshift \citep[e.g.][]{Oesch:2010,Shibuya:2015,Ma:2018}, which also explains why dust temperature increases with redshift in the $T$--$\LIR$ relation. 

Similar to those in Fig. \ref{fig:lirlpeak}, the black arrows show how galaxies at the faint/bright end move if $\fdust$ increases by a factor of 2. Dust temperature does not change at the faint end, but decreases by a factor of $\sim2^{1/6}$ at the bright end. 

\begin{figure*}
\centering
\includegraphics[width=\linewidth]{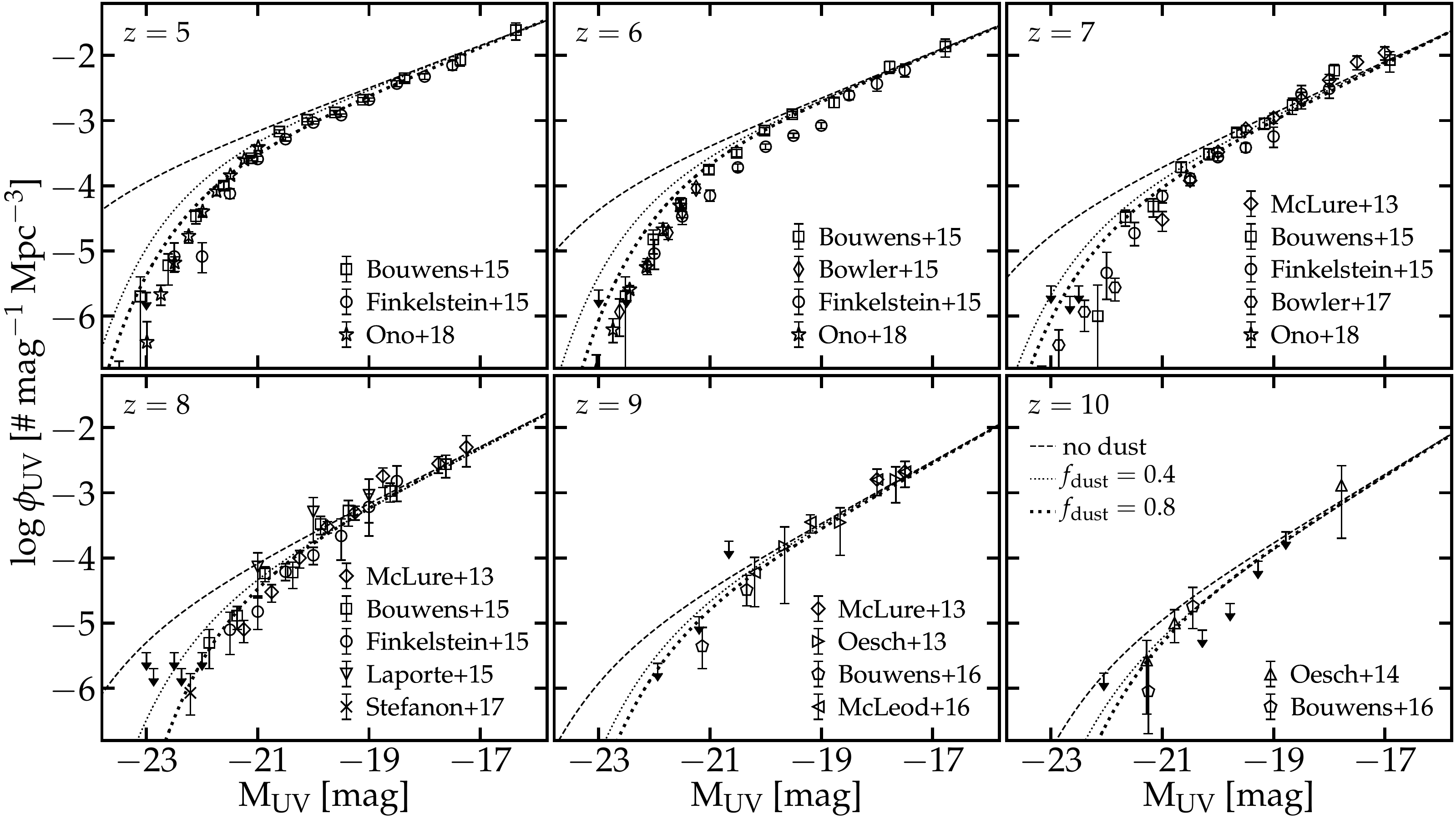}
\caption{The UVLFs from $z=5$--10. Each panel represents one redshift. The lines show the best-fit Schechter functions for the UVLFs derived from our simulated sample, with the dashed line showing the intrinsic UVLFs without dust attenuation and the thin and thick dotted lines show the post-extinction UVLFs for $\fdust=0.4$ and 0.8, respectively. We compare our results with the most up-to-date observational constraints at these redshifts from wide-field deep surveys. In all cases, the faint-end slope $\alpha$ steepens, the break magnitude ${\rm M_{UV}^{\ast}}$ increases, and the normalization $\phi_{\rm UV}^{\ast}$ decreases with redshift, which are inherited from the redshift evolution of the HMFs. The faint-end UVLFs are not strongly affected by $\fdust$. The bright-end ($\muv<-21$) UVLFs are mainly determined by dust attenuation, with ${\rm M_{UV}^{\ast}}$ increasing with $\fdust$. At $z\leq7$, the UVLFs where $\fdust=0.8$ agree better with observation than $\fdust=0.4$, although there is still a small (less than a factor of 2) discrepancy at the bright end. Such discrepancy disappears at $z\geq8$, tentatively suggesting that dust properties in $z=9$--10 galaxies are different than those in $z=5$--6 galaxies.}
\label{fig:uvlf}
\end{figure*}

\section{Results: luminosity functions and cosmic star formation rate density}
\label{sec:lf}

\subsection{The bright-end UV luminosity functions}
\label{sec:uvlf}
In this section, we construct galaxy rest-frame UVLFs at $z\geq5$ using the entire simulation sample. Fig. \ref{fig:nhalo} shows the number of halos that contain at least $10^4$ particles and have zero contamination from low-resolution particles in all 34 zoom-in regions in every 0.25\,dex bin from $\log\Mhalo=7.5$--12 at selected redshifts. There are 57 snapshots from $z=12$ to $z=5$ with 15--20\,Myr between snapshots. All halos above $\Mhalo=10^{10}\,\Msun$ and central halos above $10^{9.5}\,\Msun$ are processed with {\sc skirt}, for each of which we calculate mock images and SEDs along five lines of sight. We treat each halo snapshot and each sightline as independent `galaxies'. We do not include subhalos and satellites in this work following \citet{Ma:2018a}.

The rest-frame UV luminosity of high-redshift galaxies is usually measured in small apertures and only regions with sufficiently high surface brightness can be picked up \citep[e.g.][]{Ma:2018,Borlaff:2018}. To mimic these effects, we only include the light within an aperture of $\Rmax/3$ in projected radius to exclude satellites and diffuse starlight \citep[see e.g.][and Fig. \ref{fig:image} for examples]{Ma:2018a}.\footnote{Note that this is different from the UV luminosities in Section \ref{sec:result}, where we include all the light within $\Rvir$.} For galaxies processed with {\sc skirt}, we measure their UV luminosities directly from the mock image. For other galaxies, we project their star particles along a random sightline to produce an image, where the UV luminosity of each particle is calculated from the same stellar population synthesis models as in {\sc skirt} for consistency. Note that they are all low-mass galaxies where dust attenuation is negligible (less than 0.01\,mag seen in halos below $\Mhalo\sim10^{10}\,\Msun$).

\begin{table*}
\caption{Best-fit Schechter function (Equation \ref{eqn:schechter}) for the UVLFs derived from our simulated sample shown in Fig. \ref{fig:uvlf}. The parameters are $(\alpha, \, {\rm M_{UV}^{\ast}}, \, \phi_{\rm UV}^{\ast})$.}
\begin{tabular}{cccc}
\hline
redshift & no dust & $\fdust=0.4$ & $\fdust=0.8$ \\
\hline
$z=5$ & $(-1.81,-24.12,-4.12)$ & $(-1.81,-21.94,-3.42)$ & $(-1.90,-21.77,-3.55)$ \\
$z=6$ & $(-1.87,-23.28,-4.10)$ & $(-1.87,-21.78,-3.59)$ & $(-1.87,-21.34,-3.44)$ \\
$z=7$ & $(-1.99,-23.38,-4.57)$ & $(-2.01,-21.95,-4.08)$ & $(-2.05,-21.73,-4.09)$ \\
$z=8$ & $(-2.08,-23.07,-4.88)$ & $(-2.12,-21.66,-4.36)$ & $(-2.08,-20.97,-3.98)$ \\
$z=9$ & $(-2.18,-22.69,-5.17)$ & $(-2.17,-21.53,-4.62)$ & $(-2.20,-21.30,-4.57)$ \\
$z=10$ & $(-2.29,-21.95,-5.23)$ & $(-2.36,-21.34,-5.06)$ & $(-2.31,-20.90,-4.74)$ \\
\hline
\end{tabular}
\label{tbl:uvlf}
\end{table*}

At each redshift, we collect all `galaxies' within a $\Delta z=\pm0.5$ interval. We count the number of objects in 36 halo mass bins from $\log\Mhalo=7.5$--12 (i.e. bin width $\Delta\log\Mhalo=0.125$\,dex). On the other hand, we obtain the halo mass function (HMF) at this redshift using the public {\tt HMFcalc} code \citep{Murray:2013}, which agrees well with that directly extracted from our DM-only cosmological boxes. Every galaxy in the $i^{\rm th}$ mass bin is assigned a weight representing its abundance in the universe, $w_i = \phi_i \Delta\log M/N_i$, where $\phi_i$ is the HMF evaluated at the bin center (in Mpc$^{-3}$\,dex$^{-1}$), $\Delta\log M$ is the bin width (0.125\,dex), and $N_i$ is the number of galaxies in this bin. Next, all galaxies in the $\Delta z=\pm0.5$ redshift interval are divided in 30 equal-width bins of UV magnitude from $\muv=-24$--$-14$.\footnote{The most luminous galaxies in our sample are slightly above $\LUV=10^{12}\,\Lsun$, corresponding to $\muv=-24$. In this work, we are mainly interested in the bright-end UVLFs, so a lower limit at $\muv=-14$ is applied. We also assume that more massive halos contribute little to the UVLFs at these magnitudes, because of their low number densities in the universe as well as heavier dust obscuration in these systems.} The number density of galaxies in each $\muv$ bin is thus derived by summing over their weights. In Appendix \ref{app:lf}, we provide a detailed example about how we derive the $z=6$ UVLF from our simulated sample for interested readers. We use a \cite{Schechter:1976} function
\be
\label{eqn:schechter}
    \phi_{\rm UV} = (0.4 \, \ln10) \, \phi_{\rm UV}^{\ast} \, 10^{0.4 (\alpha+1)({\rm M_{UV}^{\ast}}-\muv)} \, e^{-10^{0.4({\rm M_{UV}^{\ast}}-\muv)}}
\ee
to fit the UVLFs derived from our simulations. We visually inspect the results to confirm that the best-fit Schechter function is always a good description for our simulated sample.

In Fig. \ref{fig:uvlf}, we present the bright-end UVLFs from $z=5$--10. Each panel shows the results at one redshift. The lines represent the best-fit Schechter functions for the UVLFs derived from our simulated sample, with the dashed lines showing the {\em intrinsic} UVLFs (without dust attenuation) and the thin and thick dotted lines showing the post-extinction UVLFs for $\fdust=0.4$ and 0.8, respectively. Again, we remind that our experiments with different $\fdust$ here is equivalent to varying the dust opacity at fixed $\fdust$. We compare our results with the most up-to-date observational constraints at these redshifts from wide-field deep surveys \citep[e.g.][symbols with errorbars]{McLure:2013,Oesch:2013,Oesch:2014,Bouwens:2015,Bouwens:2016a,Finkelstein:2015,Laporte:2015,Bowler:2017,Stefanon:2017a,Ono:2018}. We also provide the best-fit parameters of the Schechter functions for our UVLFs in Table \ref{tbl:uvlf} for reference.

In all three cases ($\fdust=0$, 0.4, and 0.8), the faint-end slope $\alpha$ becomes steeper, the break magnitude $\rm M_{UV}^{\ast}$ increases (i.e. becomes fainter), and the normalization $\phi_{\rm UV}^{\ast}$ decreases with increasing redshift. These features are primarily inherited from the redshift evolution of the HMFs. The UVLFs at $\muv>-19$ are not strongly affected by dust attenuation and the faint-end slope $\alpha$ remains unchanged with $\fdust$ at a given redshift. On the other hand, the bright-end UVLFs are determined by dust attenuation, with the break magnitude $\rm M_{UV}^{\ast}$ increasing with $\fdust$, consistent with previous results found by different authors \citep[e.g.][]{Cullen:2017,Wilkins:2017,Ma:2018a,Yung:2018}. The intrinsic UVLFs are above the observational constraints at the bright end and dust attenuation reduces the number of bright galaxies at any redshift.

At UV magnitude $\muv>-19$, the UVLFs derived from our simulations agree well with observations regardless of $\fdust$ as dust attenuation is always subdominant in this regime. For $\fdust=0.4$, the bright-end UVLFs still lie above the observational constraints at $z\leq8$. The $\fdust=0.8$ UVLFs agree better with observations, but there is still a small discrepancy (within a factor of 2) below $z=7$ at $\muv<-21$. Interestingly, such discrepancy disappears at $z\geq8$ for $\fdust=0.8$ and even the UVLFs for $\fdust=0.4$ agree well with observations at $z=9$ and 10. Although we note that the UVLFs at $z>8$ are still poorly constrained, our results here show a tentative evidence that dust properties in $z=9$--10 galaxies may be different than those in $z=5$--6 galaxies (e.g. dust fraction by mass is possibly lower at $z\geq9$). This is not unreasonable because the cosmic time at $z\geq9$ is too short for dust production from asymptotic giant branch stars in contrast to the local Universe \citep[e.g.][]{Dwek:2014}. Nonetheless, we suggest that better constraints of the bright-end UVLFs at $z>8$ with ongoing and future wide-field deep surveys can improve our understanding on dust formation and dust properties in the very early Universe in the foreseeable future. 

\begin{figure}
\centering
\includegraphics[width=\linewidth]{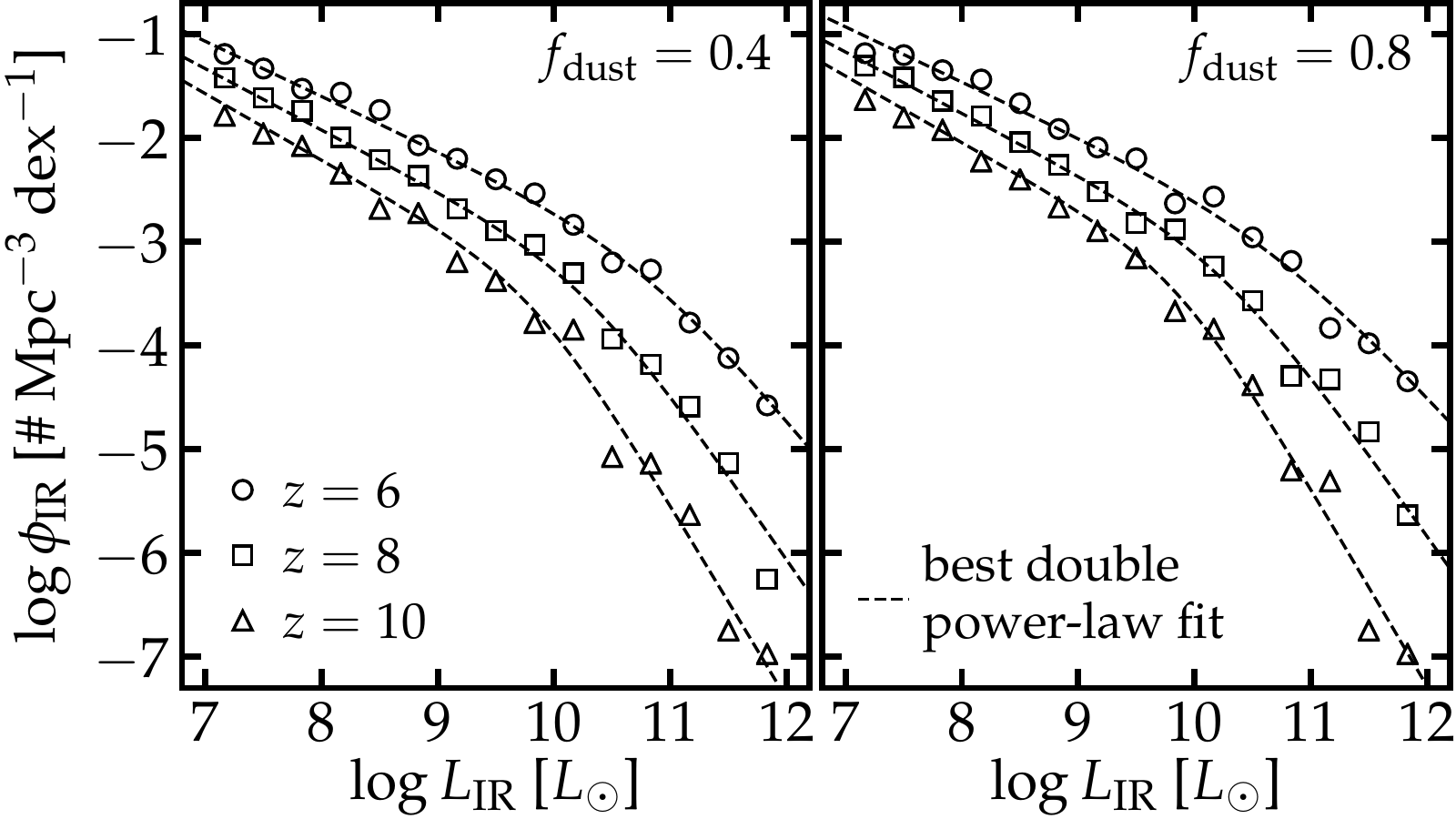}
\caption{The IRLFs at $z=6$, 8, and 10. The symbols show the results derived from our simulated sample following the method described in Section \ref{sec:irlf} for $\fdust=0.4$ (left) and 0.8 (right). The dashed lines show the best-fit redshift-dependent double power-law function (Equation \ref{eqn:irlf}).}
\label{fig:irlf}
\end{figure}

\subsection{The IR luminosity functions}
\label{sec:irlf}
In this section, we predict the bolometric IRLFs at $z=5$--10. These predictions are very useful for planning future wide-field surveys of dusty galaxies at $z\geq5$ (e.g. CSST, TolTEC/LMT) by providing a basis for estimating the number of objects one will be able to probe for a given survey volume and flux limit. Unlike rest-frame UV, the IR emission is nearly isotropic, so we do not account for line-of-sight variations, but only include each galaxy snapshot once in our analysis below. Again, at each redshift, we collect all galaxy snapshots within $\Delta z=\pm0.5$ and assign weights to halos in 36 equal-with mass bins from $\log\Mhalo=7.5$--12 as in Section \ref{sec:uvlf}. We divide all galaxies in 15 equal-width bins from $\log\LIR=7$--12 and obtain the number density of galaxies in each $\LIR$ bin by adding their weights. Note that most galaxies brighter than $\LIR\sim10^7\,\Lsun$ have been processed with {\sc skirt}. For those without dust radiative transfer calculations, their $\LIR$ are derived from $\LUV$ using Equation \ref{eqn:lirluv}. This has little effect on our results. 

In Fig. \ref{fig:irlf}, we show the derived IRLFs at $z=6$, 8, and 10 (symbols) for $\fdust=0.4$ (left) and 0.8 (right). We fit our results at integer redshifts from $z=5$--10 all together using a redshift dependent double power-law function \citep[cf.][]{Casey:2018}
\be
\label{eqn:irlf}
    \phi_{\rm IR} = \frac{\phi_{\rm IR}^{\ast}}{ \left(\frac{\LIR}{L_{\rm IR}^{\ast}}\right)^{\alpha_1} + \left(\frac{\LIR}{L_{\rm IR}^{\ast}}\right)^{\alpha_2} } ,
\ee
where $\alpha_1$, $\alpha_2$, $L_{\rm IR}^{\ast}$, and $\phi_{\rm IR}^{\ast}$ are power-law functions of $1+z$. We show the best-fit IRLFs at $z=6$, 8, and 10 in Fig. \ref{fig:irlf} (dashed lines) and the best-fit parameters are
\begin{eqnarray*}
    & \alpha_1 = 0.53 \, \left[ (1+z)/7 \right]^{0.43}, ~
    \alpha_2 = 1.37 \, \left[ (1+z)/7 \right]^{0.69}, & \\
    & L_{\rm IR}^{\ast} = 10^{10.84} \, \left[ (1+z)/7 \right]^{-4.98}, ~
    \phi_{\rm IR}^{\ast} = 10^{-3.10} \, \left[ (1+z)/7 \right]^{-1.55}, &
\end{eqnarray*}
for $\fdust=0.4$ and
\begin{eqnarray*}
    & \alpha_1 = 0.52 \, \left[ (1+z)/7 \right]^{0.43}, ~
    \alpha_2 = 1.26 \, \left[ (1+z)/7 \right]^{0.91}, & \\
    & L_{\rm IR}^{\ast} = 10^{10.80} \, \left[ (1+z)/7 \right]^{-4.60}, ~
    \phi_{\rm IR}^{\ast} = 10^{-2.94} \, \left[ (1+z)/7 \right]^{-1.59}, &
\end{eqnarray*}
for $\fdust=0.8$, respectively. The redshift-dependent double power-law function describes our results very well. Note that our simulation sample only covers up to $\LIR\sim10^{12}\,\Lsun$ and does not capture rare, most heavily obscured, extremely luminous IR galaxies (e.g. $\LIR\sim10^{13}\,\Lsun$), so our results should not be extrapolated to higher $\LIR$ without caution.

\begin{figure}
\centering
\includegraphics[width=\linewidth]{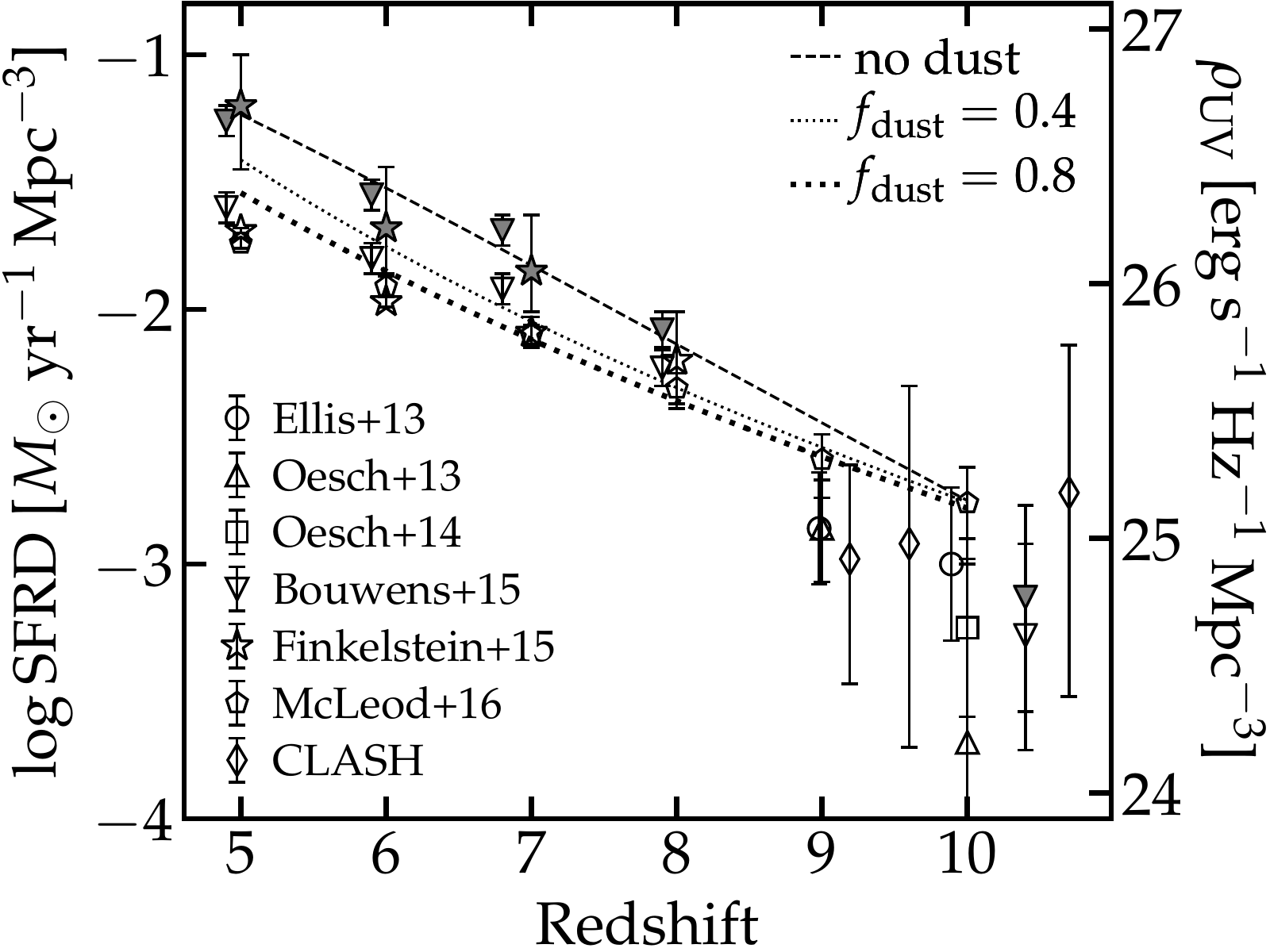}
\caption{UV luminosity density and cosmic SFRD at $z=5$--10. The lines show the results derived from our simulations. The open and filled symbols show the observational constraints in the literature with and without dust obscuration. Our results broadly agree with observations. Using $\fdust=0.4$ underestimates the obscured fraction at $z<8$, but a heavy dust attenuation is not required at higher redshifts.}
\label{fig:sfrd}
\end{figure}

\subsection{The cosmic star formation rate density}
\label{sec:sfrd}
Current observational constraints on the cosmic SFRD at $z\geq5$ are converted from rest-frame UV luminosity density using
\be
    L_{1500\,\AA} = 8.0 \times 10^{27} \, \left( \frac{\rm SFR}{\Msun\,\yr^{-1}} \right) \, {\rm erg\,s^{-1}\,Hz^{-1}}.
\ee
In this section, we calculate the dust (un)obscured UV luminosity densities at $z=5$--10 using our simulations. At each redshift, we derive the UVLF following the steps in Section \ref{sec:uvlf} and integrate the best-fit Schechter function over $\muv<-17$ (as most observational studies do) to compute the UV luminosity density at that redshift. In Fig. \ref{fig:sfrd}, we present our results for $\fdust=0$ (unobscured, dashed line), 0.4 (thin dotted line), and 0.8 (thick dotted line) and compare with observational constraints (symbols with errorbars; e.g. \citealt{Ellis:2013,Oesch:2013,Oesch:2014,Bouwens:2015,Finkelstein:2015,McLeod:2016}; and CLASH detections from \citealt{Zheng:2012,Coe:2013,Bouwens:2014}). The open (filled) symbols show the (un)obscured results, respectively.

Our predicted unobscured UV luminosity density agrees fairly well with current observational constraints. Similar to the case with UVLFs in Section \ref{sec:uvlf}, we find our $\fdust=0.8$ results agree better with the observed dust obscured UV luminosity density than those using $\fdust=0.4$, as the latter underestimate the obscured fraction of the UV light at $z<8$. At higher redshifts, the difference between $\fdust=0.4$ and 0.8 becomes much smaller and thus a heavy dust attenuation is no longer required at $z\geq8$.

\begin{figure}
\centering
\includegraphics[width=\linewidth]{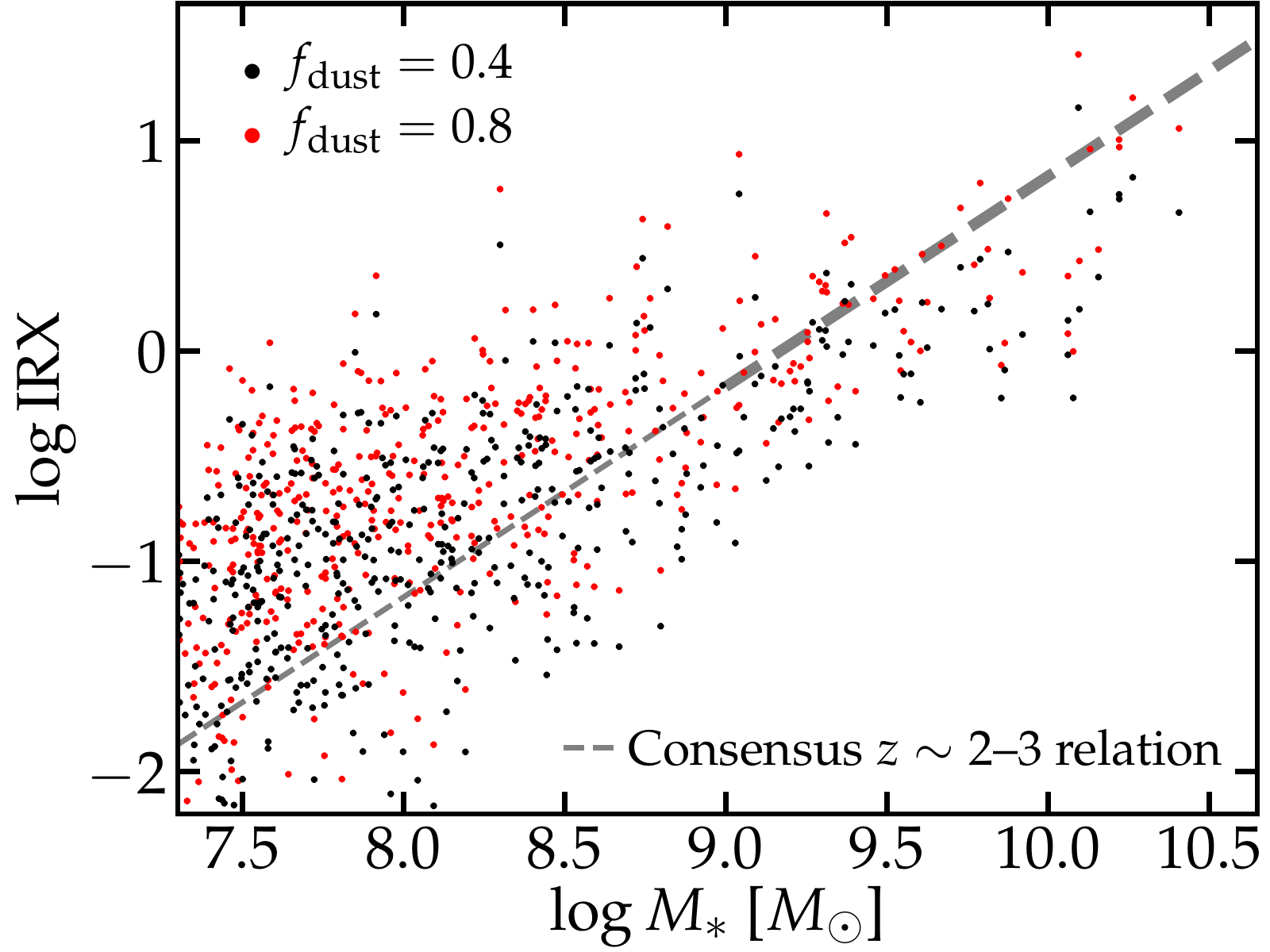}
\caption{The IRX--stellar mass relation. Each point represents one sightline of a galaxy snapshot at integer redshift from $z=5$--12, for $\fdust=0.4$ (black) and 0.8 (red). The grey dashed line shows the consensus relation for $z\sim2$--3 galaxies \citep{Reddy:2010,Whitaker:2014,Alvarez-Marquez:2016}. The thick segment represents the mass range where current observational constraints are available, while the thin segment shows the interpolation of the relation to lower masses. For individual galaxies, IRX is typically smaller by $\sim0.3$\,dex for $\fdust=0.4$ than $\fdust=0.8$ (each galaxy appears as a pair of red and black points at the same $\Ms$). This suggests that the IRX--stellar mass relation can be used to constrain $\fdust$.}
\label{fig:irxms}
\end{figure}

\section{Discussion}
\label{sec:discussion}

\subsection{Strategies for probing dusty galaxies at $z\geq5$}
\label{sec:strategy}
In this paper, we present a broad spectrum of predictions on dust attenuation and emission in high-redshift galaxies that can be tested and motivate future observations. First of all, we argue that current data cannot completely rule out the existence of an IRX--$\buv$ relation in $z\geq5$ galaxies that is consistent with local relations. The UV-continuum slope of a galaxy may still be a good indicator of dust attenuation. Where galaxies lie on the IRX--$\buv$ relation is predominantly determined by the shape of the dust extinction curve in the UV, which reflects the dust composition. Better constraints of the IRX--$\buv$ relation at $z\geq5$ with the possibility of constraining the attenuation law in the rest-frame optical in the future can help understand dust formation history and the evolution of dust properties across cosmic time.

We predict that dust temperatures can be much higher in $z\geq5$ galaxies than in low-redshift galaxies, a consequence of high sSFR (luminosity per unit dust mass) and/or high SFR surface densities (intensity of radiation on dust grains) in high-redshift galaxies. This can be tested using multi-band observations of dust emission from a sample of intrinsically bright, dust obscured galaxies at $z\geq5$. It would be interesting to select targets based on their UV-continuum slopes, but the observations must be deeper than current surveys in the (sub)mm, as the flux densities at the R--J tail are reduced by a factor of a few (cf. Fig. \ref{fig:sed}). Having coverage on at least one band at a wavelength shorter than rest-frame $\lp$ is critical for measuring dust temperature and bolometric dust luminosity, which will be achievable with the OST (wavelength coverage from 5--600\,$\mu$m). This also helps us understand whether galaxies with red UV slopes but low apparent IRX are real or just because their bolometric IR luminosities are underestimated from single-wavelength data due to the presence of warmer dust.

We find that the IRX--$\buv$ relation does not depend on dust fraction or dust-to-gas ratio. On the other hand, as shown in Section \ref{sec:uvlf}, the shape of the bright-end UVLFs is sensitive to dust fraction. We therefore propose that the bright-end UVLFs can be combined with IRLFs, the IRX--$\buv$ relation, and other observables at long wavelengths as a new method to infer dust properties in $z\geq5$ galaxies in a statistical sense. Ongoing and future observations with the {\it Hubble Space Telescope} and the {\it James Webb Space Telescope} in the rest-frame UV as well as current and next-generation radio telescopes (e.g. ALMA, CSST, ngVLA, SPICA, OST) probing dust emission in high-redshift galaxies are very promising to this end. 

In Fig. \ref{fig:irxms}, we also examine the IRX--stellar mass relation for our simulated sample, using $\fdust=0.4$ (black points) and 0.8 (red points), respectively. Each galaxy appears as a pair of red and black points at the same $\Ms$. This relation is independent of redshift and we only show one sightline for each galaxy snapshot at integer redshift from $z=5$--12. The IRX--stellar mass relation depends on $\fdust$, with IRX decreasing roughly by $\sim0.3$\,dex for individual galaxies if $\fdust$ drops from 0.8 to 0.4. The grey dashed line shows the consensus $z\sim2$--3 IRX--stellar mass relation \citep[e.g.][]{Reddy:2010,Whitaker:2014,Alvarez-Marquez:2016}. The thicker part represents relatively massive galaxies (i.e. $\Ms>10^9\,\Msun$) for which current observational constraints are available at $z\sim2$--3, while the thinner part represents its interpolation to lower masses. Our simulations broadly agree with this relation, in line with the results in \cite{Bouwens:2016} where they find that the inferred IRX--stellar mass relation of typical $\sim L^{
\ast}$ galaxies at $z\sim4$--10 from 1.2\,mm ALMA-HUDF deep survey is consistent with the $z\sim2$--3 relation if dust temperature increases with redshift to $\sim44$--50\,K at $z\sim6$. Our results suggest that better constraints on the IRX--stellar mass relation at $z\geq5$ can also be used to infer dust fraction in $z\geq5$ galaxies in addition to bright-end UVLFs.

\subsection{Limitations of this work}
\label{sec:limit}
In this work, we include starlight as the sole source heating the dust and only study the `intrinsic' dust emission and dust temperature. We note that heating from the CMB can play a significant role at $z\geq5$, when the CMB temperature starts to become comparable to the dust temperature. Following the argument in \cite{da-Cunha:2013} for single-temperature dust, the CMB first heats the dust to a higher temperature 
\be
\label{eqn:cmb}
    T_{\rm dust,\,with\,CMB} = \left( T_{\rm dust,\,intrinsic}^{4+\beta} + T_{\rm CMB}^{4+\beta} \right)^{\frac{1}{4+\beta}},
\ee
where $\beta=2$ is the dust emissivity index and $T_{\rm CMB}$ is the CMB temperature at the redshift of interest. Second, the CMB serves as a background which the dust emission from high-redshift galaxies is measured against. Subtracting this background reduces the observed flux by a factor of $1-B_{\nu}(T_{\rm CMB})/B_{\nu}(T_{\rm dust,\,with\,CMB})$ at a given frequency. In general, galaxies with higher intrinsic dust temperatures are less affected than those with primarily cold dust. The net effect is stronger at longer wavelengths than at shorter wavelengths.

We exclude the CMB in this paper on purpose for two reasons. First, empirical models of number counts or luminosity functions of high-redshift galaxies in the IR and (sub-)mm often start from intrinsic dust emission and then convert to observed flux following \cite{da-Cunha:2013} as summarized above \citep[e.g.][]{Bouwens:2016,Casey:2018}. Therefore, it is important to understand the dust SEDs and dust temperatures in $z\geq5$ galaxies without the CMB. Second, the effects of the CMB are more complicated in reality given the broad distribution of dust temperature in individual galaxies. Warm dust close to young stars is barely affected, while the diffuse dust at much lower temperature mostly becomes invisible. It is not clear which dust temperature is applicable to Equation \ref{eqn:cmb} and by what fraction the observed flux is reduced in different regions of a galaxy. In a future study, we will investigate how CMB heating affects the observed far-IR flux from our simulated galaxies using full radiative transfer calculations where we include the CMB as an extra source that produces a uniform radiation field with a black-body spectrum.

Our dust radiative transfer calculations assume a fixed dust composition and dust-to-metal ratio everywhere in a galaxy as well as in all galaxies. In reality, dust composition may vary in different regions of a galaxy, as seen in the MW where the extinction curve varies between lines of sight. Moreover, the dust-to-metal ratio in cold, dense gas is presumably higher than that in warm, diffuse gas, because dust growth is more efficient and the grains are less likely to be destroyed in cold, dense gas. This may further enlarge the discrepancy of dust attenuation between young stars just born in dense clouds and relatively older stars preferentially living in more diffuse gas, leading to a dramatic effect on the galaxy-averaged attenuation law. Furthermore, in the local Universe, it has been suggested that the dust-to-metal ratio decreases at low metallicity \citep[below $\sim0.1\,\Zsun$, e.g.][]{Remy-Ruyer:2014}. However, it is not clear if it is also the case at high redshifts, given that the ISM conditions are very different from those at lower redshifts. Given that the dust properties in high-redshift galaxies are still poorly constrained because there are only a small set of data available so far, we adopt the simplest treatments in this work to qualitatively predict what to expect for future observations. More sophisticated treatments of dust physics will be necessary in the future if new data suggest so.

Last but not least, our simulation sample only includes normal star-forming galaxies that are typically discovered in current deep surveys in the rest-frame UV. The most massive galaxies in our sample have stellar mass $\sim10^{10.5}\,\Msun$ and bolometric IR luminosity $\sim10^{12}\,\Lsun$. We do not yet simulate more massive, heavily obscured, and luminous systems (e.g. $\Ms\sim10^{11}\,\Msun$, $\LIR\sim10^{13}\,\Lsun$) at these redshifts at comparably high resolution, like the extremely luminous DSFGs detected by SPT \citep[e.g.][]{Strandet:2016}. They are relatively rare objects that may involve major mergers of two massive galaxies or rapidly accreting supermassive black holes. It is not clear where such galaxies lie on the IRX--$\buv$ (stellar mass) relation and whether they have higher dust temperatures than lower-redshift galaxies at similar luminosities. There is no guarantee that our predictions in this paper still hold for more massive and luminous systems.

\section{Conclusions}
\label{sec:conclusion}
In this work, we utilize a suite of 34 cosmological zoom-in simulations that consist of thousands of sufficiently resolved halos spanning a halo mass range $\Mhalo\sim10^8$--$10^{12}\,\Msun$ with stellar mass up to $\sim10^{10.5}\,\Msun$ and {\em intrinsic} UV luminosity up to $\sim10^{12}\,\Lsun$ ($\muv\sim-24$) at $z\geq5$. These simulations use the FIRE-2 models of the multi-phase ISM, star formation, and stellar feedback. With a mass relation of $7000\,\Msun$ or better and typical spatial resolution in dense gas of 1\,pc, these simulations explicitly resolve star formation in dense birth clouds and feedback destroying these clouds. We post-processing all halos above $\Mhalo=10^{10}\,\Msun$ and central halos above $10^{9.5}\,\Msun$ in our sample using the three-dimensional Monte Carlo dust radiative transfer code {\sc skirt} to study dust attenuation, dust emission, and dust temperature in high-redshift galaxies. Our calculations assume a SMC-like dust composition from \cite{Weingartner:2001} and a constant dust-to-metal ratio $\fdust$ in all gas below $10^6$\,K (no dust in hotter gas). We fix dust composition and opacity but experiment with different $\fdust$, which accounts for uncertainties of dust opacity and dust-to-metal ratio in a single parameter. We do not adopt any models for sub-resolution dust distribution but instead process the simulations directly. Our main findings include the following.

(i) Dust geometry is clumpy and patchy. The young stars emitting most of the UV photons are usually concentrated in the central region of the galaxy, but dust is distributed on much larger spatial scales. Dust scatters UV light to an extended distribution at relatively low surface brightness, which contributes a non-negligible fraction of the escaped UV flux (Fig. \ref{fig:image}).

(ii) Our sample shows a tight relationship between IR excess (IRX) and UV-continuum slope ($\buv$), consistent with the SMC IRX--$\buv$ relation, despite the patchy dust geometry in our simulations. Galaxies at higher redshifts tend to move slightly to bluer $\buv$ at fixed IRX due to their younger stellar population. Viewing the same galaxy from different sightlines gives the same IRX--$\buv$ relation as the entire sample (Fig. \ref{fig:irx}, top panel).

(iii) The IRX--$\buv$ relation does {\em not} depend on the normalization of the attenuation law (represented by different $\fdust$). However, it {\em does} depend on the shape of the extinction curve (Fig. \ref{fig:irx}, bottom panel), which reflects the dust composition.

(iv) Our simulations produce an IRX--stellar mass relation in broad agreement with the consensus relation established for $z\sim2$--3 galaxies. This relation depends on $\fdust$, with IRX decreasing by $\sim0.3$\,dex if $\fdust$ drops from 0.8 to 0.4 (Fig. \ref{fig:irxms}).

(v) There is a positive correlation between bolometric IR luminosity $\LIR$ and {\em intrinsic} UV luminosity $\LUVi$, which can be described by a broken power-law function (Equation \ref{eqn:lirluv}). At the bright end, $\LIR$ changes little with $\fdust$ (the optically-thick limit), while at the faint end, $\LIR$ is proportional to $\fdust$ (the optically-thin limit). The $\LIR$--$\LUVi$ relation does not depend on redshift (Fig. \ref{fig:lirluv}, left panel).

(vi) The scatter in the $\LIR$--$\LUVi$ relation is not driven by dust mass, average dust column density, nor dust density, although all three quantities are redshift-dependent. This suggests that dust luminosity is mainly determined by dust covering fraction. There is a secondary correlation between $\LIR$ and the SFR averaged over the past 10\,Myr (i.e. the amount of stars younger than 10\,Myr) at a given $\LUVi$, because young stars are more heavily obscured than relatively older stars (Fig. \ref{fig:lirluv}, right panel).

(vii) The correlation between $\LIR$ and $\SFR_{\rm 10\,Myr}$ for the entire sample can be well described by a power-law function. The scatter of this relation is driven by $\LUVi$. This further confirms the differential obscuration between stars younger and older than 10\,Myr (Fig. \ref{fig:lirsfr}). Note that $\LUVi$ and $\rm SFR_{10\,Myr}$ are not fully degenerated.

(viii) Our simulated sample shows an anti-correlation between the peak wavelength $\lp$ of dust emission (in terms of $L_{\nu}$) and $\LIR$. However, the $\lp$--$\LIR$ relation for $z\geq5$ galaxies shows a large offset from the observed relation at lower redshifts, with $\lp$ moving toward shorter wavelengths by a factor of 2 at a given $\LIR$ (Fig. \ref{fig:lirlpeak}), suggesting that dust is on average warmer in high-redshift galaxies. 

(ix) The dust SEDs are far from an optically-thin MBB function. At $z=6$, the flux densities at ALMA Band 6 (observed-frame 1.2\,mm) of our simulated galaxies are comparable to MBB spectra with $T\sim45$--60\,K at the same $\LIR$ (Fig. \ref{fig:sed}). The low detection rate of dust continuum at $z\geq5$ compared to what inferred from the UV slopes is likely due to higher dust temperatures in these galaxies.

(x) We predict that dust temperature correlates positively with both sSFR (approximately dust luminosity per unit mass) and SFR surface density (intensity of the interstellar radiation). Both correlations are independent of redshift. At fixed $\LIR$, dust temperature increases with redshift from $z=5$--12 (Fig. \ref{fig:Tcomp}), because galaxies at higher redshifts tend to have higher sSFR and more compact SFR. Dust temperature does not change significantly with $\fdust$ at the faint end, but increases by a factor of $2^{1/6}=1.12$ if $\fdust$ increases by a factor of 2.

(xi) Using the entire simulation sample, we derive the UVLFs from $z=5$--10. The bright-end UVLFs are largely determined by dust attenuation. By comparing our results with most up-to-date observational constraints, we find tentative evidence that dust properties are likely evolving from $z=10$ to $z=5$ (Fig. \ref{fig:uvlf}, Table \ref{tbl:uvlf}). We suggest that better measurements of the bright-end UVLFs at $z>8$ with future observations provide a powerful probe of dust physics in the very early Universe.

(xii) We predict the bolometric IRLFs up to $\LIR\sim10^{12}\,\Lsun$ at $z=5$--10, which can be described by a redshift-dependent double power-law function (Equation \ref{eqn:irlf}, Fig. \ref{fig:irlf}).

(xiii) We derive dust (un)obscured UV luminosity density and cosmic SFRD at $z=5$--10 from the UVLFs. Our results are broadly consistent with observational constraints in the literature (Fig. \ref{fig:sfrd}).

\section*{Acknowledgement}
We thank our referee, Maarten Baes, for helpful suggestions and for pointing out our mistakes in an earlier version of this paper in describing the technical details of the {\sc skirt} code.
The simulations used in this paper were run on XSEDE computational resources (allocations TG-AST120025, TG-AST130039, TG-AST140023, and TG-AST140064). 
CMC thanks the University of Texas at Austin College of Natural Sciences, NSF grants AST-1714528, AST-1814034, and a 2019 Cottrell Scholar Award for support from the Research Corporation for Science Advancement.
PFH was supported by an Alfred P. Sloan Research Fellowship, NASA ATP Grant NNX14AH35G, and NSF Collaborative Research Grant \#1411920 and CAREER grant \#1455342.
CAFG was supported by NSF through grants AST-1517491, AST-1715216, and CAREER award AST-1652522; by NASA through grant 17-ATP17-0067; by STScI through grant HST-AR-14562.001; and by a Cottrell Scholar Award from the Research Corporation for Science Advancement.
EQ was supported by NASA ATP grant 12-APT12-0183, a Simons Investigator award from the Simons Foundation, and the David and Lucile Packard Foundation.
RF acknowledges financial support from the Swiss National Science Foundation (grant no 157591).
DK was supported by NSF grant AST-1412153, funds from the University of California, San Diego, and a Cottrell Scholar Award from the Research Corporation for Science Advancement. 
The Flatiron Institute is supported by the Simons Foundation.

\bibliography{library}

\appendix

\section{Convergence tests}
\label{app:conv}

\begin{figure}
\centering
\includegraphics[width=\linewidth]{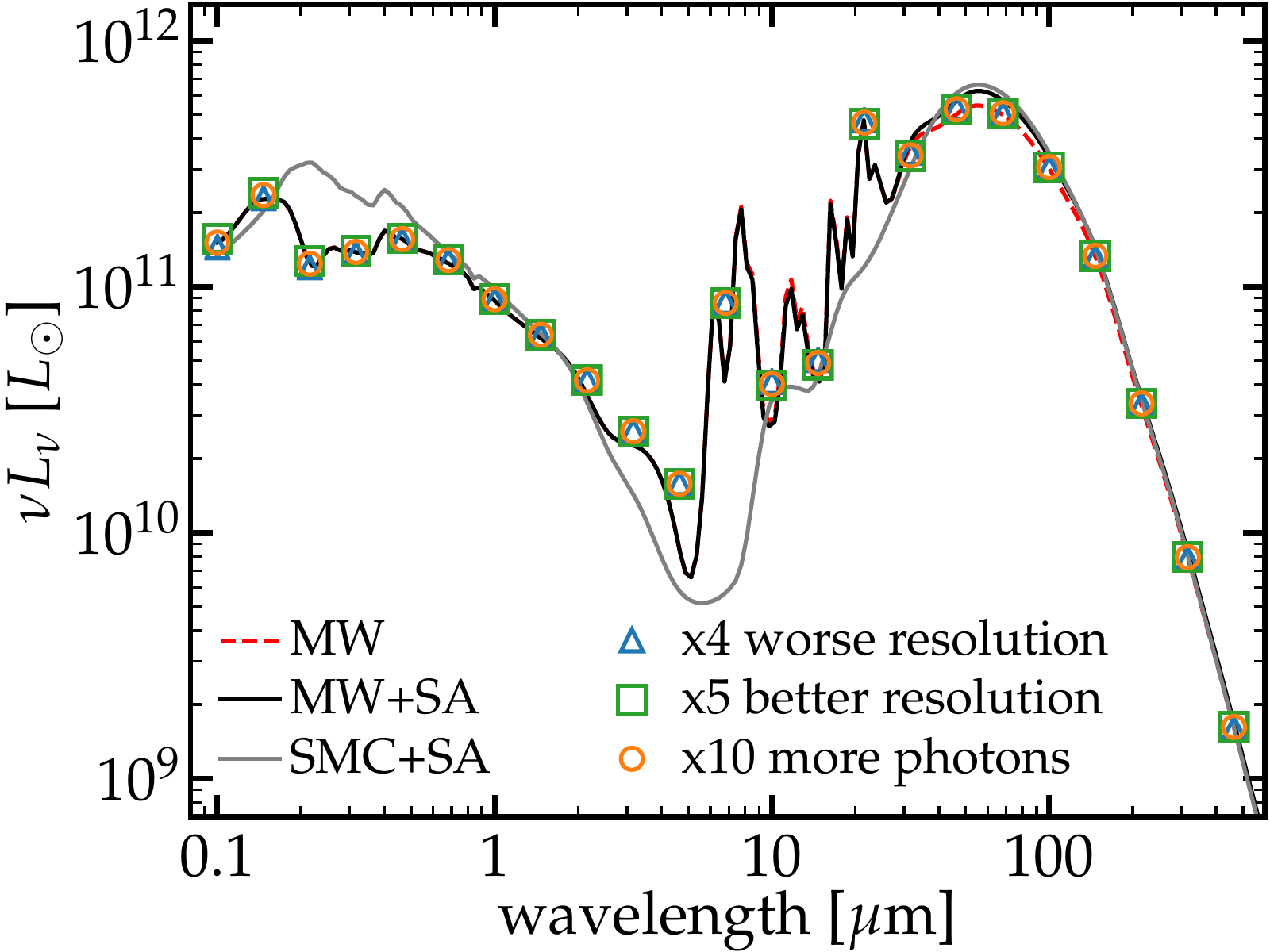}
\caption{Galaxy SEDs of galaxy z5m12b using different parameters for dust radiative transfer calculations. The grey solid line represents our default model. The black solid line uses the MW-type dust model from \citet{Weingartner:2001}. The red dashed line uses the same parameters as the black line, except that dust self-absorption is ignored. The SEDs using MW-type dust and SMC-type dust models mainly differ (1) in the mid-IR (rest-frame 3--25\,$\mu$m) due to PAH emission and (2) the near-UV absorption feature because of the 2175\,{\AA} bump in the MW-like extinction curve. The symbols show that our default choices of dust grid resolution and number of photon packets ensure excellent resolution.}
\label{fig:conv}
\end{figure}

In Section \ref{sec:rt}, we describe our choices for the dust transfer calculations. We adopt the built-in octree dust grid in {\sc skirt}, which is constructed from gas particles and adaptively refines the high-density region until the dust mass in a cell is less than $10^{-6}$ of the total dust mass in the domain. We use $10^6$ photon packets at each of the 90 wavelengths equally spaced in logarithmic scale from 0.08--1000\,$\mu$m. We use the SMC dust grain size distribution model from \cite{Weingartner:2001}, which consist of carbonaceous and silicate grains but no PAH. Dust self-absorption is included.

In Fig. \ref{fig:conv}, we compare the SEDs of galaxy z5m12b using different parameters in the dust radiative transfer calculations. The grey solid line represents our default choices. The black solid line uses the same grid resolution and number of photon packets, but the MW-like grain size distribution from \cite{Weingartner:2001}, which include a PAH component. The red dashed line uses the same parameters as the black solid line, but without dust self-absorption. The mid-IR SED (6--25\,$\mu$m) differs significantly between MW-type dust and SMC-type dust due to PAH emission. For MW-type dust model, there is a strong absorption feature in the near-UV caused by the 2175\,{\AA} bump in the extinction curve, making the rest-frame UV slope $\buv$ always negative, so we do not use MW-type dust as our default choice. Nevertheless, at rest-frame $\lambda>30\,\mu$m, both models give nearly identical results. We note that dust self-absorption is important around the peak wavelength of dust emission.

The symbols show resolution tests using MW-type dust model without dust self-absorption on a 25-point wavelength grid. Orange circles show the results using $10^7$ photon packets per wavelength. Blue triangles show the results using a factor of 4 worse resolution for the dust grid (i.e. each cell has a maximum dust mass equals to $4\times10^{-6}$ of the total dust mass in the domain). Green square show the results using a factor of 5 better resolution for the dust grid (the maximum dust mass in each cell is $2\times10^{-7}$ of the total dust mass). These calculations agree precisely well with each other, suggesting that our default choices for grid resolution and the number of photon packets (the red dashed line) ensure excellent convergence.

\begin{figure}
\centering
\includegraphics[width=\linewidth]{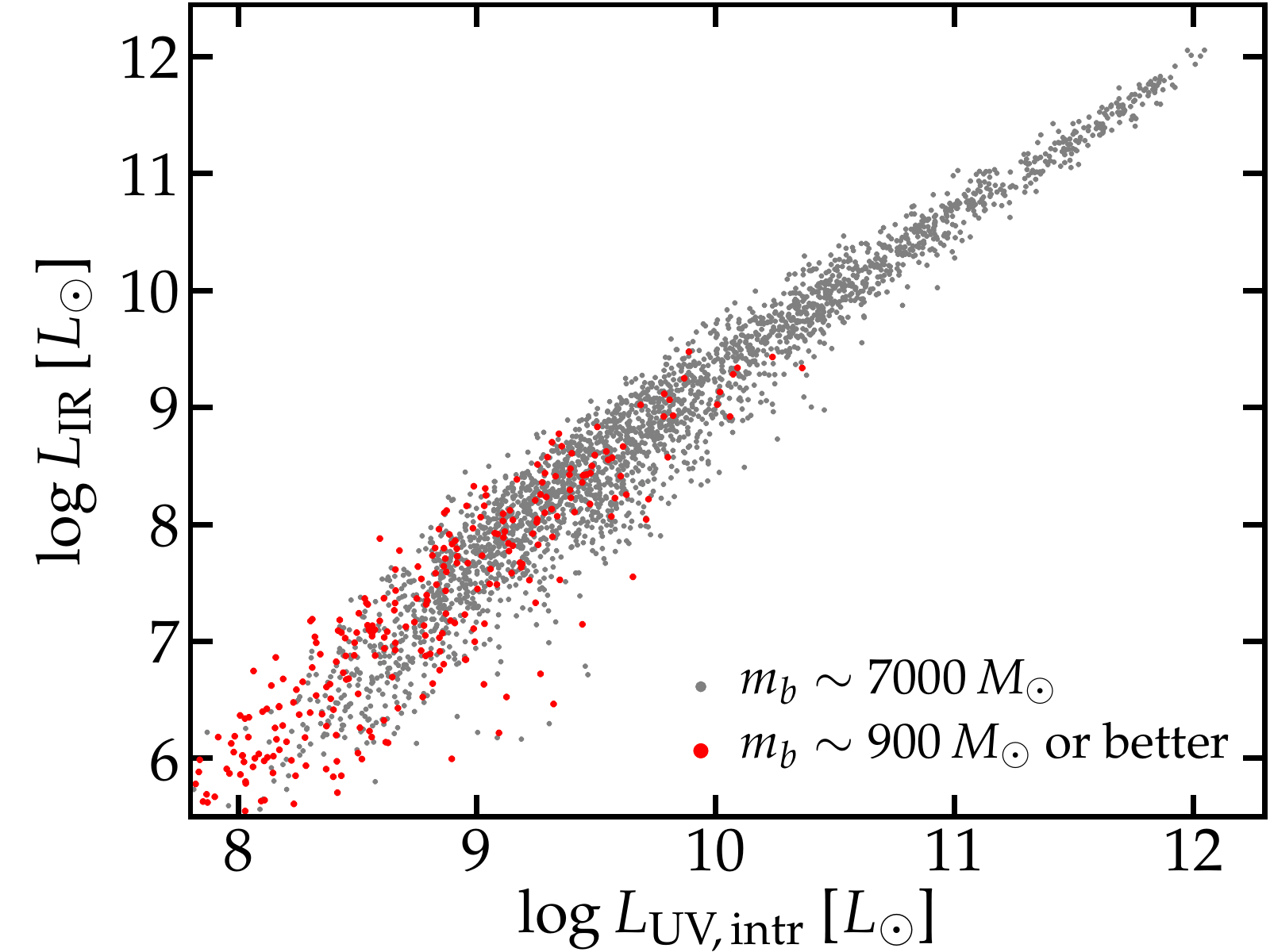}
\caption{The $\LIR$--$\LUVi$ relation for $\fdust=0.4$, same as the left panel in Fig. \ref{fig:lirluv}. Galaxies simulated at mass resolution $\mb\sim7000\,\Msun$ are shown as grey points, while those simulated at $\mb\sim900\,\Msun$ resolution or better are shown as red points. Our results are not sensitive to the resolution of the simulations, at least in the range where the two subsamples overlap.}
\label{fig:restest}
\end{figure}

In Fig. \ref{fig:restest}, we show the same $\LIR$--$\LUVi$ relation as in the left panel of Fig. \ref{fig:lirluv}, but separate galaxies simulated at mass resolution $\mb\sim7000\,\Msun$ and at resolution $\mb\sim900\,\Msun$ or better with grey and red points, respectively. The two subsamples of our simulated galaxies overlap at halo mass $\Mhalo\lesssim10^{10.5}\,\Msun$. There is no significant difference on the $\LIR$--$\LUVi$ relation between the two subsamples. We also explicitly check all the results in this paper and find that none of our conclusions is sensitive to the resolution of our simulations, at least in the mass range where the two subsamples overlap.

\section{How we derive UVLFs from the simulated sample}
\label{app:lf}

\begin{figure*}
\centering
\includegraphics[width=\linewidth]{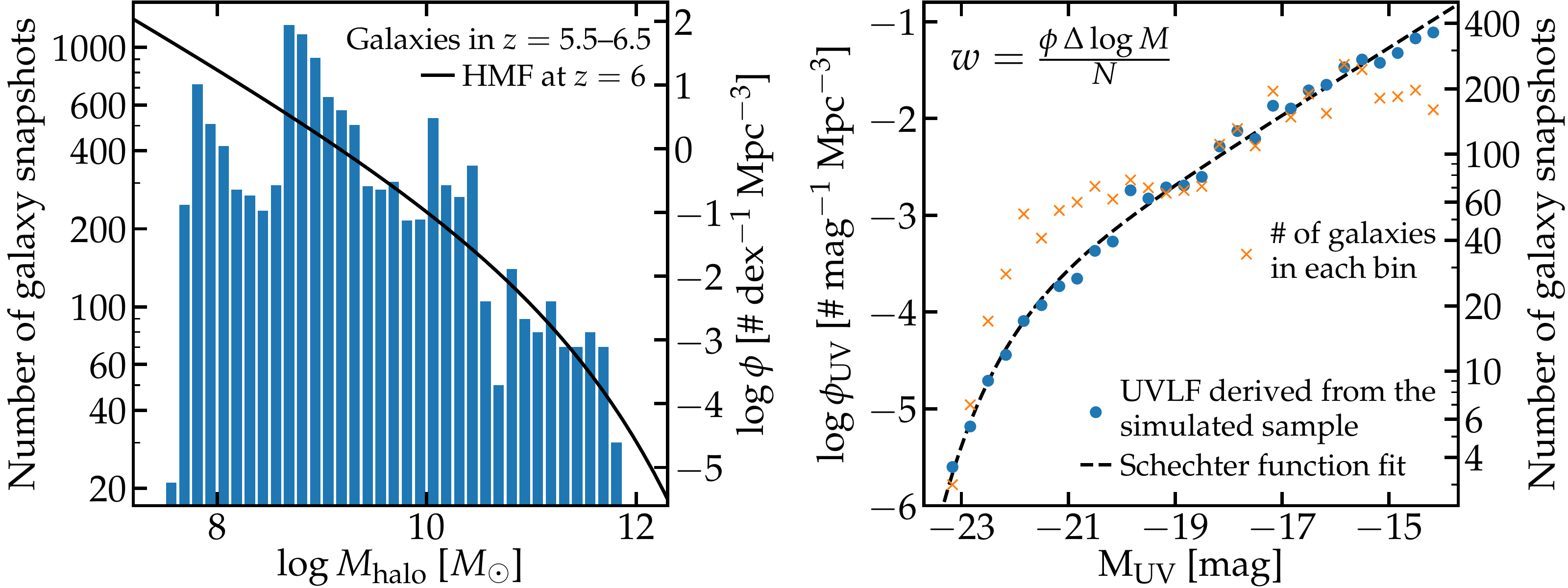}
\caption{{\em Left}: Number of galaxies in 36 bins of halo mass for all galaxies in $5.5<z<6.5$ in our sample from $\log\Mhalo=7.5$--12 (histogram). The black solid line shows the $z=6$ HMF calculated from the {\tt HMFcalc} code. {\em Right}: Number of galaxies in 30 bins of $\muv$ from $-24$ to $-14$\,mag (orange symbols). The blue circles show the $z=6$ UVLF derived from our simulated sample, by summing the weight over all galaxies in each $\muv$ bin and dividing it by the bin width. Using a smaller number of $\muv$ bins reduces the noise but does not affect our results. The black dashed line shows the best-fit Schechter function of the $z=6$ UVLF (Equation \ref{eqn:schechter}).}
\label{fig:weight}
\end{figure*}

In Section \ref{sec:uvlf}, we briefly describe the methods to construct UVLFs using our simulated sample. Here we walk through the steps in detail for interested readers using an example, the $z=6$ UVLF. The histogram in the left panel of Fig. \ref{fig:weight} shows the number of `galaxies' in 36 equal-width bins of $\log\Mhalo$ from 7.5--12 (i.e. bin width 0.125\,dex) in $z=5.5$--6.5. Note that we treat those in different snapshots and the same galaxy viewed along different sightlines as different objects. The black solid line in the left panel shows the HMF at $z=6$ obtained from {\tt HMFcalc} code. Every galaxy in the same halo mass bin is given the same weight representing its number density in the universe, $w=\phi\,\Delta\log M/N$, where $\phi$ is the HMF evaluated at the bin center, $\Delta\log M=0.125\,\dex$ is the bin width, and $N$ is the number of objects in this bin.

In the right panel, we show the number of galaxies in 30 bins of $\muv$ from $-24$\,mag to $-14$\,mag (orange cross). The blue circles show the total weight of galaxies in each $\muv$ bin, divided by the bin width (i.e. $1/3$\,mag). This is thus the UVLF derived from our simulated sample. Note that there is small noise in the result, which is caused by the fluctuations in the number of galaxies in each $\muv$ bin. Using a smaller number of bins will reduce the noise, but do not affect our results significantly. The black dashed line shows the best-fit Schechter function of the $z=6$ UVLF (Equation \ref{eqn:schechter}). We visually inspect and confirm that a Schechter function is always a good description of the UVLF derived from our sample at any redshift. 



\label{lastpage}

\end{document}